\documentclass[fleqn,usenatbib,useAMS]{mnras}

\usepackage[T1]{fontenc}
\usepackage{ae,aecompl}

\usepackage{graphicx}	
\usepackage{amsmath}	
\usepackage{amssymb}	
\usepackage{pdflscape}
\usepackage{times,txfonts}
\usepackage{hyperref}

\usepackage{verbatim}
\usepackage{gensymb} 
\usepackage{bm}         
\usepackage{adjustbox}
\usepackage[super]{nth}
\usepackage{scalerel}[2014/03/10]
\usepackage{stackengine}
\usepackage{ulem}
\usepackage{listings}

\def\url#1{\expandafter\string\csname #1\endcasname}
\newlength{\colwidth}
\newcommand{\kpc}                      {\,{\rm kpc}}

\newcommand{\pkpc}                      {\,{\rm pkpc}}

\newcommand{\Msun}                    {\,{\rm M}_\odot}

\newcommand{\kms}                      {\,\,{\rm km}\,\,{\rm s}^{-1}}

\newcommand{\gadget}         {\textsc{gadget3}}
\newcommand{\subfind}        {\textsc{subfind}}

\def\lsim{\mathrel{\lower0.6ex\hbox{$\buildrel {\textstyle <}
 \over {\scriptstyle \sim}$}}}

\title[Morphology and kinematics of EAGLE galaxies]{The relationship between the morphology and kinematics of galaxies and its dependence on dark matter halo structure in EAGLE}

\author[A. C. R. Thob et al.]{Adrien C. R. Thob,$^{1}$\thanks{E-mail: A.Thob@2015.ljmu.ac.uk (LJMU)} 
Robert A. Crain,$^{1}$
Ian G. McCarthy,$^{1}$
Matthieu Schaller,$^{2}$
\newauthor 
Claudia D. P. Lagos,$^{3,4}$
Joop Schaye,$^{2}$
Geert Jan J. Talens,$^{2}$
Philip A. James,$^{1}$
\newauthor 
Tom Theuns,$^{5}$
and Richard G. Bower,$^{5}$
\\
$^{1}$Astrophysics Research Institute, Liverpool John Moores University, 146 Brownlow Hill, Liverpool L3 5RF, UK\\
$^{2}$Leiden Observatory, Leiden University, P.O. Box 9513, 2300 RA Leiden, the Netherlands\\
$^{3}$International Centre for Radio Astronomy Research (ICRAR), M468, University of Western Australia, 35 Stirling Hwy, Crawley, WA 6009, Australia\\
$^{4}$ARC Centre of Excellence for All-sky Astrophysics in 3 Dimensions (ASTRO 3D)\\
$^{5}$Institute for Computational Cosmology, Durham University, South Road, Durham DH1 3LE, UK}

\date{Accepted XXX. Received YYY; in original form ZZZ}

\pubyear{2018}

\begin{document}
\label{firstpage}
\pagerange{\pageref{firstpage}--\pageref{lastpage}}
\maketitle

\begin{abstract}
We investigate the connection between the morphology and internal kinematics of the stellar component of central galaxies with mass $M_\star > {10}^{9.5} \,\Msun$ in the EAGLE simulations. We compare several kinematic diagnostics commonly used to describe simulated galaxies, and find good consistency between them. We model the structure of galaxies as ellipsoids and quantify their morphology via the ratios of their principal axes. We show that the differentiation of blue star-forming and red quiescent galaxies using morphological diagnostics can be achieved with similar efficacy to the use of kinematical diagnostics, but only if one is able to measure both the flattening and the triaxiality of the galaxy. Flattened oblate galaxies exhibit greater rotational support than their spheroidal counterparts, but there is significant scatter in the relationship between morphological and kinematical diagnostics, such that kinematically-similar galaxies can exhibit a broad range of morphologies. The scatter in the relationship between the flattening and the ratio of the rotation and dispersion velocities ($v/\sigma$) correlates strongly with the anisotropy of the stellar velocity dispersion: at fixed $v/\sigma$, flatter galaxies exhibit greater dispersion in the plane defined by the intermediate and major axes than along the minor axis, indicating that the morphology of simulated galaxies is influenced significantly by the structure of their velocity dispersion. The simulations reveal that this anisotropy correlates with the intrinsic morphology  of the galaxy's inner dark matter halo, i.e. the halo's morphology that emerges in the absence of dissipative baryonic physics. This implies the existence of a causal relationship between the morphologies of galaxies and that of their host dark matter haloes. 
\end{abstract}




\section{Introduction}
\label{sec:introduction}

The morphology and internal kinematics of galaxies are fundamental characteristics, both of which have an established tradition as a means to classify the galaxy population and infer aspects of its evolution over cosmic time. The two properties are closely related, with flattened, disky galaxies primarily supported by rotation, whilst spheroidal or elliptical galaxies exhibit greater dispersion support \citep[for recent observational findings see][]{van_de_Sande_et_al_17,van_de_Sande_et_al_18b,Graham_et_al_18}. Moreover, it is well established that both quantities correlate broadly with other properties, for example mass \citep[e.g.][]{Dressler_80,Baldry_et_al_06, Kelvin_et_al_14}, colour \citep[e.g.][]{Blanton_et_al_03,Driver_et_al_06} and star formation rate \citep{Kennicutt_83,Kauffmann_White_and_Guiderdoni_93}, indicating that they encode information relating to the formation history of galaxies. In particular, the recognition that the specific angular momentum of stars is markedly higher in late-type galaxies than in early-type counterparts \citep{Fall_and_White_83,Romanowsky_and_Fall_12,Fall_and_Romanowsky_18} led to the development of analytic galaxy evolution models in which the latter more readily dissipate their angular momentum throughout their assembly \citep[e.g.][]{Fall_and_Efstathiou_80,Mo_Mao_and_White_98}, for example as a consequence of a more intense merger history. 

The relatively recent advent of large surveys conducted with wide-field integral field spectrographs has enabled the compilation of large and diverse samples of galaxies in the local Universe with well-characterised morphological and kinematical properties \citep[e.g.][]{de_Zeeuw_et_al_02,Cappellari_et_al_11,Croom_et_al_12,Sanchez_et_al_12,Ma_et_al_14,Bundy_et_al_15}. One of the prime outcomes of these endeavours is the demonstration that there is a not a simple mapping between galaxy morphology and internal kinematics, particularly within the family of early-type galaxies for which the kinematics are not generally dominated by rotation \citep[for a recent review see][]{Cappellari_16}. Early-type galaxies with similar morphologies are found to exhibit a diversity of kinematic properties, indicating that kinematic diagnostics may yield a more fundamental means of classifying galaxies than purely morphological descriptions \citep[e.g.][]{Emsellem_et_al_07,Krajnovic_et_al_13,Cortese_et_al_16,Graham_et_al_18}. Similarly, \citet{Foster_et_al_17} recently showed that the morphologies of kinematically-selected galaxies are clearly correlated with the degree of rotational support, but with a large degree of scatter. Analytic modelling of galaxies using the tensor virial theorem indicates that this diversity stems from differing degrees of anisotropy in the stellar velocity dispersion \citep[e.g.][]{Binney_1976}, but the origin of the diversity in the inferred anisotropy remains unclear.

Several families of cosmological simulations of galaxy formation have recently emerged that reproduce key characteristics and scaling relations exhibited by the observed galaxy population \citep[see e.g.][]{Vogelsberger_et_al_14b,S15,Kaviraj_et_al_17,Pillepich_et_al_18a}. Such simulations evolve the dark matter and baryonic components self-consistently from cosmologically-motivated initial conditions, and the morphological and kinematical properties of galaxies emerge in response to this assembly. Crucially, the current generation of state-of-the-art cosmological simulations do not suffer from `catastrophic overcooling' \citep{Katz_and_Gunn_91,Navarro_and_Steinmetz_97,Crain_et_al_09}, a failure to adequately regulate the inflow of gas onto galaxies, which results in the formation of a galaxy population that is generally too old, too massive, too compact, and too dispersion-supported. This success is in part due to improvements in the numerical treatment of hydrodynamical processes, but more importantly is due to the implementation of feedback treatments that effectively regulate and quench star formation \citep[e.g.][]{Okamoto_et_al_05,Scannapieco_et_al_08,Governato_et_al_09,Dalla_Vecchia_and_Schaye_12,Scannapieco_et_al_12,C15} and preferentially eject low angular momentum gas from the interstellar medium \citep[e.g.][]{Sommer_Larsen_et_al_10,Brook_et_al_11,Agertz_et_al_11}.

Numerical simulations of large cosmic volumes therefore afford the opportunity to examine the relationship between the morphological and kinematical properties of a well-sampled population of galaxies, the origin of scatter about such a relationship, and the connection between these properties and other observables such as mass, star formation rate and photometric colour. The markedly improved realism of the current generation of state-of-the-art simulations engenders greater confidence in conclusions drawn from their analysis.

In this study we examine the relationship between the morphology and internal kinematics of galaxies formed in the EAGLE simulations of galaxy formation \citep{S15,C15}. We compare the kinematic properties of EAGLE galaxies with quantitative morphological diagnostics, enabling a rigorous examination of the relationship between the two properties and their connection to other observables. The simulation also enables us to investigate the origin of scatter about the relation between the two properties. We have added the morphological and kinematical diagnostics computed for this study to the public EAGLE database, enabling their use by the wider community. This work complements several related studies of the morphological and/or kinematical properties of EAGLE simulations, such as \citet{Correa_et_al_17}, who show that the kinematic properties of EAGLE galaxies can be used as a qualitative proxy for their visual morphology and that this morphology correlates closely with a galaxy's location in the colour-mass diagram; \citet{Lagos_et_al_18b}, who investigated the role of mass, environment and mergers in the formation of `slow rotators'; \citet{Clauwens_et_al_18}, who identified three phases of morphological evolution in galaxies, primarily as a function of their stellar mass; and \citet{Trayford_et_al_19}, who explored the emergence of the Hubble `tuning fork' sequence. 

This paper is structured as follows. We discuss our numerical methods in Section \ref{sec:methods}, providing a brief summary of the simulations and galaxy finding algorithms, and detailed descriptions of our techniques for characterising the morphology and internal kinematics of simulated galaxies. In Section \ref{sec:morphokinem} we first examine the morphology and internal kinematics of EAGLE galaxies, and the relationship of these quantities with the location of the galaxies in the colour-mass diagram, before turning to the relationship between the morphology and internal kinematics. In Section \ref{sec:origin_anisotropy} we consider the origin of scatter about the relation between the two diagnostics. We summarise and discuss our findings in Section \ref{sec:summary}. In Appendix \ref{sec:convergence} we present a brief test of the influence of numerical resolution on the relationship between morphology and kinematics. In Appendix \ref{sec:virial} we present a brief analytical derivation of the relationship between morphology, kinematics and the anisotropy of the velocity dispersion from the tensor virial theorem. Finally, in Appendix \ref{sec:public} we present an explanation of how to access the morphological and kinematical diagnostics computed here from the public EAGLE database.

\section{Numerical methods}
\label{sec:methods}

In this section we present an overview of the EAGLE simulations, including a concise description of the most relevant subgrid physics implementations and methods for identifying galaxies and their host haloes. Summaries of these methods are included in many papers that focus on analyses of EAGLE \citep[e.g.][]{Crain_et_al_17} but we retain key details here for completeness, so readers familiar with the simulations may wish to skip to Section \ref{sec:shape_params}. We subsequently introduce the diagnostics used to characterise the morphology and kinematics of our simulated galaxies; software enabling the reader to compute these diagnostic quantities can be obtained from the public repository at http://github.com/athob/morphokinematics.

\subsection{Simulations and subgrid physics}
\label{sec:simulations}

EAGLE \citep[Evolution and Assembly of GaLaxies and their Environments;][]{S15,C15} is a suite of hydrodynamical simulations of the formation, assembly and evolution of galaxies in the $\Lambda$CDM cosmogony, whose data have been publicly released \citep{McAlpine_et_al_16}. The EAGLE simulations are particularly attractive for our purposes, because the model was explicitly calibrated to reproduce the stellar masses and sizes of the present-day galaxy population. Comparison with multi-epoch observations highlights that the stellar masses \citep{Furlong_et_al_15}, sizes \citep{Furlong_et_al_17} and angular momenta \citep[e.g.][]{Swinbank_et_al_17} of EAGLE's galaxy population also evolve in a realistic fashion.

The EAGLE simulations adopt cosmological parameters from \citet{Planck_2014_paperI_short}, namely $\Omega_0 = 0.307$, $\Omega_{\rm b} = 0.04825$, $\Omega_\Lambda= 0.693$, $\sigma_8 = 0.8288$, $n_{\rm s} = 0.9611$, $h = 0.6777$, $Y = 0.248$. They were evolved using a version of the $N$-body TreePM smoothed particle hydrodynamics (SPH) code \gadget, last described by \citet{Springel_05}. This version incorporates modifications to the hydrodynamics algorithm and the time-stepping criteria, and includes a series of subgrid routines that govern, in a phenomenological fashion, physical processes that act on scales below the resolution limit of the simulations.

At `standard resolution', the EAGLE simulations have particle masses corresponding to a volume of side $L = 100$ comoving Mpc (hereafter cMpc) realized with $2 \times 1504^3$ particles (an equal number of baryonic and dark matter particles), such that the initial gas particle mass is $m_{\rm g} = 1.81 \times 10^6 \Msun$, and the mass of dark matter particles is $m_{\rm dm} = 9.70 \times 10^6 \Msun$. The Plummer-equivalent gravitational softening length is fixed in comoving units to 1/25 of the mean interparticle separation (2.66 comoving kpc, hereafter ckpc) until $z = 2.8$, and in proper units (0.70 proper kpc, hereafter pkpc) thereafter. The standard-resolution simulations marginally resolve the Jeans scales at the density threshold for star formation in the warm and diffuse photoionised ISM. High-resolution simulations adopt particle masses and softening lengths that are smaller by factors of 8 and 2, respectively.

The updates to the hydrodynamics algorithm, which are detailed in Appendix A of \citet{S15}, comprise the pressure-entropy formulation of SPH of \citet{Hopkins_13}, the \citet{Cullen_and_Dehnen_10} artificial viscosity switch, an artificial conduction switch similar to that proposed by \citet{Price_08}, the use of the \citet{Wendland_95} $C^2$ smoothing kernel, and the \citet{Durier_and_Dalla_Vecchia_12} time-step limiter. The influence of these developments on the galaxy population realised by the simulations is explored in the study of \citet{Schaller_et_al_15b_short}.

Gas particles denser than a metallicity-dependent density threshold for star formation \citep{Schaye_04} become eligible for conversion into stellar particles. The probability of stochastic conversion is dependent on the gas particle's pressure \citep{Schaye_and_Dalla_Vecchia_08}. Supermassive black holes (BHs) are seeded in haloes identified by a friends-of-friends (FoF) algorithm run periodically during the simulation, and they grow by gas accretion and mergers with other BHs \citep[][]{Springel_Di_Matteo_and_Hernquist_05,Booth_and_Schaye_09,S15}. The gas accretion rate onto BHs is influenced by the angular momentum of gas close to the BH \citep[see][]{Rosas_Guevara_et_al_15} and cannot exceed the Eddington limit.

Feedback associated with the evolution of massive stars (`stellar feedback') and the growth of BHs (`AGN feedback') is also implemented stochastically \citep{Dalla_Vecchia_and_Schaye_12}. The efficiency of stellar feedback is a function of the local density and metallicity of each newly-formed stellar particle; the dependence of the feedback efficiency on these properties was calibrated to ensure that the simulations reproduce the present-day galaxy stellar mass function, and produce disc galaxies with realistic sizes \citep{C15}. The efficiency of AGN feedback was calibrated such that the simulations reproduce the relationship between the stellar masses of galaxies and the masses of their central BHs, at the present day. 

The mass of stellar particles is $\sim 10^6\,{\rm M}_\odot$, so each can be modelled as a simple stellar population (SSP). We assume the initial mass function (IMF) of stars of the form proposed by \citet{Chabrier_03}, with masses $0.1-100\,{\rm M}_\odot$. The return of mass and nucleosynthesised metals from stars to the interstellar medium (ISM) is implemented as per \citet{Wiersma_et_al_09}; this scheme follows the abundances of the 11 elements most important for radiative cooling and photoheating (H, He, C, N, O, Ne, Mg, Si, S, Ca and Fe), using nucleosynthetic yields for massive stars, Type Ia SNe, Type II SNe and the AGB phase. Element-by-element radiative cooling and heating of gas is implemented as per \citet{Wiersma_Schaye_and_Smith_09}, assuming the to be optically thin and in ionisation equilibrium with the cosmic microwave and metagalactic UV backgrounds. 

The simulations lack the resolution to model the cold, dense phase of the ISM explicitly. They thus impose a temperature floor, $T_{\rm eos}(\rho)$, which prevents the spurious fragmentation of star-forming gas. The floor takes the form of an equation of state $P_{\rm eos} \propto \rho^{4/3}$ normalised so $T_{\rm eos} = 8000 {\rm K}$ at $n_{\rm H} = 0.1 {\rm cm}^{-3}$. The temperature of star-forming gas therefore reflects the effective pressure of the ISM, rather than its actual temperature. Since the Jeans length of gas on the temperature floor is $\sim 1\,{\rm pkpc}$, a drawback of its use is that it suppresses the formation of gaseous discs with vertical scale heights much shorter than this scale. However, as recently shown by \citet{Benitez_Llambay_et_al_18}, the primary cause of the thickening of non-self-gravitating discs in EAGLE is likely to be turbulent pressure support stemming from the gas accretion and energy injection from feedback, and the influence of the latter is likely to be artificially high. We comment further on the implications of this thickening for our study in Section \ref{sec:shape_params}.

Our analyses focus primarily on the simulation of the largest volume, Ref-L100N1504. To facilitate convergence testing (presented in Appendix \ref{sec:convergence}), we also examine the high-resolution Recal-L025N0752 simulation. The feedback efficiency parameters adopted by this model were recalibrated to ensure reproduction of the calibration diagnostics at high resolution \citep[see][]{C15}.

\subsection{Identifying and characterizing galaxies}
\label{sec:subhalo}

We consider galaxies as the stellar component of gravitationally self-bound structures. The latter are identified using the \subfind{} algorithm \citep{Springel_et_al_01,Dolag_et_al_09}, applied to haloes identified using the friends-of-friends algorithm (FoF). The substructure, or `subhalo', hosting the particle with the lowest gravitational potential in each halo is defined as the `central' subhalo, with all others considered as satellite subhaloes, which may host satellite galaxies. The coordinate of this particle defines the centre of the galaxy, about which is computed the spherical overdensity mass, $M_{200}$, for the adopted enclosed density contrast of 200 times the critical density. When aggregating the stellar properties of galaxies, we consider all stellar particles residing within a 3D spherical aperture of radius $30\pkpc$ centered on the galaxy's potential minimum; as shown by \citet{S15}, this yields stellar masses comparable to those recovered within a projected circular aperture of the Petrosian radius.  

To suppress environmental influences on the morphology and kinematics of galaxies, we focus exclusively on central galaxies. Since the characterisation of these properties also requires particularly good particle sampling, we require that galaxies have a present-day stellar mass $M_\star > 10^{9.5}\,\Msun$, which corresponds to a minimum of $\simeq 1700$ stellar particles. We further exclude from consideration galaxies with a resolved satellite subhalo (i.e. comprised of 20 or more particles of any type) whose mass is at least 1 percent of the central galaxy's total mass, and whose potential minimum resides within the $30\pkpc$ aperture. These selection criteria are satisfied by $4155$ present-day central galaxies in Ref-L100N1504.

\subsection{Characterising galaxy morphology with shape parameters}
\label{sec:shape_params}

As discussed in Section \ref{sec:simulations} and demonstrated by \citet{Trayford_et_al_17}, disc galaxies in the EAGLE simulations are more vertically-extended than their counterparts in nature. We therefore opt against performing a detailed structural decomposition to characterise the galaxies' morphologies, such as might be achieved by applying automated multi-component profiling algorithms \citep[see e.g.][]{Simard_98,Peng_et_al_02,Robotham_et_al_17} to mock images of the galaxies. Instead, we obtain a quantitative description of the galaxies' structures by modelling the spatial distribution of their stars with an ellipsoid, characterised by the flattening ($\epsilon$) and triaxiality ($T$) parameters. These are defined as:
\begin{eqnarray}\label{eq:shapeparameters}
\epsilon = 1 - \frac{c}{a} \, ,  \, \, \, {\rm and} \, \, \,
T = \frac{a^2 - b^2 }{a^2 - c^2 } ,
\end{eqnarray}
where $a$, $b$, and $c$ are the moduli of the major, intermediate and minor axes, respectively. For spherical haloes, $\epsilon=0$ and $T$ is undefined, whilst low and high values of $T$ correspond to oblate and prolate ellipsoids, respectively. Clearly, these diagnostics are poor descriptors of systems that deviate strongly from axisymmetry but, as noted by \citet{Trayford_et_al_19}, such galaxies are rare within the present-day galaxy population. The axis lengths are defined by the square root of the eigenvalues, $\lambda_i$ (for $i=0,1,2$), of a matrix that describes the galaxy's 3-dimensional mass distribution. The simplest choice is the tensor of the quadrupole moments of the mass distribution\footnote{As noted elsewhere, the mass distribution tensor is often referred to as the moment of inertia tensor, since the two share common eigenvectors.} \citep[see e.g.][]{DEFW85,Cole_and_Lacey_96,Bett_12}, defined as:
\begin{equation}\label{eq:inertiatensor}
\mathcal{M}_{ij}= \frac{\sum_{p} m_p r_{p,i} r_{p,j}}{\sum_{p} m_p} \, ,
\end{equation}
where the sums run over all particles comprising the structure, i.e. with $r_p<30\pkpc$, $r_{p,i}$ denotes the component $i$ (with $i,j=0,1,2$) of the coordinate vector of particle $p$, and $m_p$ is the particle's mass. However, we opt to use an iterative form of the reduced inertia tensor \citep[see e.g.][]{Dubinski_and_Carlberg_91,Bett_12,Schneider_et_al_12}. The use of an iterative scheme is advantageous in cases where the morphology of the object can deviate significantly from that of the initial particle selection, as is the case here for flattened galaxies. The reduced form of the tensor mitigates the influence of structural features in the outskirts of galaxies by down-weighting the contribution of particles farther from their centre, i.e. with larger ellipsoidal radius, $\tilde{r}_p$ (eq. \ref{eq:eq_vol}):
\begin{equation}\label{eq:redinertiatensor}
\mathcal{M}^r_{ij}= \frac{\sum_{p} \frac{m_p}{\tilde{r}_p^2} r_{p,i} r_{p,j}}{\sum_{p} \frac{m_p}{\tilde{r}_p^2}}.
\end{equation}
In the first iteration, all stellar particles comprising the galaxy (those within a spherical aperture of $r=30\pkpc$) are considered, yielding an initial estimate of the axis lengths ($a,b,c$). Stellar particles enclosed by the ellipsoid of equal volume described by the axis ratios:    
\begin{equation}\label{eq:eq_vol}
\tilde{r}_p^2 \equiv r_{p,a}^2 + \frac{r_{p,b}^2}{(b/a)^2} + \frac{r_{p,c}^2}{(c/a)^2} \leq \left ( \frac{a^2}{b c}\right )^{2/3} (30 \pkpc)^2,
\end{equation}
are then identified, where the quantities $r_{p,a}$, $r_{p,b}$ and $r_{p,c}$ are the distances projected along the directions defined by the eigenvectors calculated in the previous iteration, and the axis lengths recomputed from this set. This process continues until the fractional change of both of the ratios $c/a$ and $b/a$ converges to $< 1$ percent. Such convergence is generally achieved within 8-10 iterations, and the resulting median lengths of the aperture's major axis for galaxies of $\epsilon \simeq (0.2,0.5,0.8)$ are $a=(34,39,50)\pkpc$.
	
\subsection{Characterising galaxy kinematics}
\label{sec:kin_diag}

Several diagnostic quantities are frequently used to characterise the kinematic properties of simulated galaxies. We briefly describe five of the most commonly-adopted diagnostics below, and assess the consistency between them in Section \ref{sec:compare_kinematics}. In all cases, coordinates are computed in the frame centered on the galaxy's potential minimum, and velocities in the frame defined by the mean velocity of star particles within $30\pkpc$ of this centre. Unlike the calculation of the shape parameters, for which we consider particles within an iteratively-defined ellipsoidal aperture, the particle-based kinematic diagnostics described here are computed using stellar particles within a spherical aperture of $r=30 \pkpc$, for consistency with the existing literature.\\

\noindent \textit{Fraction of counter-rotating stars}: The mass fraction of stars that are rotationally-supported (which can be considered the `disc' mass fraction) is a simple and intuitive kinematic diagnostic. A popular means of estimating the disc fraction is to assume that the bulge component has no net angular momentum, and hence its mass can be estimated as twice the mass of stars that are counter-rotating with respect to the galaxy \citep[e.g.][]{Crain_et_al_10,McCarthy_et_al_12a,Clauwens_et_al_18}. We therefore consider the disc-to-total mass fraction, $D/T$, to be the remainder when the bulge-to-total mass fraction, $B/T$ is subtracted:
\begin{equation}
\frac{D}{T} = 1 - \frac{B}{T} = 1 - 2 \frac{1}{M_\star} \sum_{i,L_{{\rm z},i}<0} {m_i},
\label{eq:DTratio}
\end{equation}
where the sum is over all counter-rotating ($L_{{\rm z},i}<0$) stellar particles within $30\pkpc$, $m_i$ is the mass of each stellar particle and $L_{{\rm z},i}$ is the component of its angular momentum projected along the rotation axis, where the latter is the unit vector parallel to the total angular momentum vector of all stellar particles with $r < 30\pkpc$.\\

\noindent \textit{Rotational kinetic energy}: the parameter $\kappa_{\rm co}$ specifies the fraction of a particle's total kinetic energy, $K$, that is invested in co-rotation $K_{\rm co}^{\rm rot}$ \citep{Correa_et_al_17}: 
\begin{equation}
\kappa_{\rm co} = \frac{K^{\rm rot}_{\rm co}}{K} = \frac{1}{K} \sum_{i,L_{{\rm z},i}>0} {\frac{1}{2} m_i {\left ( \frac{L_{{\rm z},i}}{m_i R_i} \right )}^2},
\label{eq:kappaco}
\end{equation}
where the sum is over all co-rotating ($L_{{\rm z},i}>0$) stellar particles within $30\pkpc$, and $R_i$ is the 2-dimensional radius in the plane normal to the rotation axis. The total kinetic energy in the centre of mass frame is $K = \sum_i \frac{1}{2} m_i v_i^2$, again summing over all stellar particles within $30\pkpc$. 

\citet{Correa_et_al_17} used this diagnostic to characterise the kinematics of EAGLE galaxies, and found that dividing the population about a threshold in $\kappa_{\rm co}$ provides a means of separating the `blue cloud' ($\kappa_{\rm co} > 0.4$) of disky star-forming galaxies from the `red sequence' ($\kappa_{\rm co} < 0.4$) of spheroidal passive galaxies in the galaxy colour - stellar mass diagram. As those authors discussed, eq. \ref{eq:kappaco} differs slightly from the usual definition of $\kappa$ \citep{Sales_et_al_10}, insofar that only \textit{corotating} particles contribute to the numerator. This results in a better measure of the contribution of rotation to the kinematics of the galaxy, since the majority of counter-rotating particles are found within the bulge component. \\

\noindent \textit{Spin parameter}: we use the measurements of the mass-weighted stellar spin parameter, $\lambda_\star$, computed for EAGLE galaxies in a similar manner to the calculation of luminosity-weighted stellar spin parameters presented by \citet{Lagos_et_al_18b}. We create datacubes similar to those recovered by integral field spectroscopy, by projecting stellar particles onto a 2-dimensional grid to create a stellar mass-weighted velocity distribution for each pixel. We fit a Gaussian function to this distribution, defining the rotation velocity as that at which the Gaussian peaks, and the velocity dispersion as the square root of the variance, and arrive at the spin via:
\begin{equation}
\lambda_\star = \frac{\sum_{i}{ m_i r_i v_i} }{\sum_{i}{ m_i r_i \sqrt{ v_i^2 + \sigma_i^2} }},
\label{eq:lambdaR}
\end{equation}
where $m_i$ is the total stellar mass of the pixel $i$, $v_i$ is its line-of-sight velocity, $\sigma_i$ is its (1-dimensional) line-of-sight velocity dispersion, and $r_i$ is the pixel's 2-dimensional galactocentric radius. The sum runs over pixels enclosed within the 2-dimensional projected stellar half-mass radius, $r_{\star,1/2}$. We compute spin measurements from maps in which the galaxies are oriented edge-on with respect to the spin vector. We note that this observationally-motivated definition of the spin parameter differs from the classical definition \citep[see e.g.][]{Bullock_et_al_01}.\\ 

\noindent \textit{Orbital circularity}: the parameter\footnote{We use the symbol $\xi$ to denote the orbital circularity rather than the more commonly adopted $\epsilon$ to avoid confusion with the flattening parameter, $\epsilon$, defined in Section \ref{sec:shape_params}.} $\xi$ \citep[see e.g.][]{Abadi_et_al_03b,Zavala_et_al_16} specifies the circularity of a particle's orbit by comparing its angular momentum to the value it would have if on a circular orbit with the same binding energy:
\begin{equation}
\xi_i = \frac{j_{{\rm z},i}}{j_{\rm circ} (E_i)},
\label{eq:orbital}
\end{equation}
where $j_{{\rm z},i}$ is the particle specific angular momentum projected
along the rotation axis, and $j_{\rm circ}(E_i)$ is the specific angular momentum corresponding to a circular orbit with the same binding energy $E_i$. We estimate the latter as the maximum value of $j_{\rm z}$ for all stellar particles within $30\pkpc$ and $E < E_i$. Positive (negative) values of $\xi$ correspond to co-rotation (counter-rotation). 

An advantage of this method is that it can be used to assign particles to bulge and disc components, thus enabling a kinematically-defined structural decomposition. However, to enable a simple comparison with other kinematic diagnostics, we assign to each galaxy the median of the $\xi$ values, $\overline{\xi}$, exhibited by star particles within $30\pkpc$ of the galaxy centre.\\

\noindent \textit{Ratio of rotation and dispersion velocities}: the ratio of rotation and dispersion velocities is often used as a kinematical diagnostic since, as noted in the discussion of the spin parameter, both the rotation velocity and the velocity dispersion can be estimated from spectroscopic observations of galaxies \citep{van_de_Sande_et_al_17}. We adopt a cylindrical coordinate frame ($r$,$z$,$\theta$), with the $z$-axis parallel to the total angular momentum of stellar particles within $30\pkpc$ and the azimuthal angle increasing in the direction of net rotation, and equate the rotation velocity of each galaxy, $V_{\rm rot}$, to the absolute value of the `mass-weighted median' of the tangential velocities, $v_{\theta,i}$, of its stellar particles. We compute mass-weighted medians of variables by identifying the value that equally divides the weights, i.e. we construct the cumulative distribution of weights from rank-ordered values of the variable in question, and interpolate to the value that corresponds to the `half-weight' point. Since we weight by particle masses, this is equivalently the `half-mass value'.

To connect with observational measurements of the dispersion, which necessarily recover an estimate of the line-of-sight velocity dispersion, we seek the velocity dispersion in the `disc plane', i.e. the plane normal to the $z$-axis, which we denote $\sigma_0$. The latter represents all remaining motion within the disc plane following the subtraction of the ordered co-rotational component, which can in principle comprise coherent streaming with respect to the centre-of-velocity frame, randomly-oriented elliptical orbits, or circular orbits that are mis-aligned with the disc, including counter-rotation. This component can be computed from the tensor virial equations: for an axisymmetric system of stellar particles in a Cartesian frame and rotating about the $z$-axis, $2T_{xx} + \Pi_{xx} + W_{xx} = 0$, with $T_{xx}=T_{yy}$ and $T_{ij}=0$ for $i \neq j$, and similarly for both $\Pi$ and $W$. Here, $W$ is the potential energy and $T$ and $\Pi$ are the contributions to the kinetic energy tensor, $K$, from ordered and disordered motion, respectively, such that $K_{ij}=T_{ij}+\frac{1}{2}\Pi_{ij}$. \citet{Binney_and_Tremaine_87} show that $2T_{xx} = \frac{1}{2}M_\star V_{\rm los}^2$ (assuming rotation about the $z$-axis is the only streaming motion) and $\Pi_{xx}=M_\star \sigma_{\rm los}^2$, where $V_{\rm los}$ and $\sigma_{\rm los}$ are the line-of-sight rotation velocity and velocity dispersion, respectively. 

Since we seek the velocity dispersion in the disc plane rather than along the line-of-sight, we exploit that the disc plane and vertical contributions are separable, i.e. $2T_{zz} + \Pi_{zz} + W_{zz} = 0$ also, and use
\begin{equation}
K - K_{zz} = \frac{1}{2}M_\star V_{\rm rot}^2 + M_\star \sigma_0^2, 
\label{eq:v_over_sig}
\end{equation}
which can be rearranged and solved for $\sigma_0$, as $K$ and $K_{zz}$ are the total and vertical kinetic energies of the system of stellar particles. 

The disordered motion, $\Pi$, can also be separated into its components along the vertical axis ($M_\star \sigma_z^2$) and in the disc plane ($M_\star \sigma_0^2$); these are related via the parameter, $\delta$, which describes the anisotropy of the galaxy's velocity dispersion:
\begin{equation}
\delta = 1 - \left( \frac{\sigma_z}{\sigma_0} \right)^2.
\label{eq:anisotropy_param}
\end{equation}
Values of $\delta > 0$ indicate that the velocity dispersion is primarily contributed by disordered motion in the disc plane, i.e. that defined by the intermediate and major axes, rather than disordered motion in the direction of the minor axis. A more complete derivation of these equations is presented in Appendix \ref{sec:virial}.

\subsubsection{A brief comparison of kinematical diagnostics}
\label{sec:compare_kinematics}

\begin{figure}
\adjustbox{trim={.01\width} {.093\height} {.04\width} {.087\height},clip}%
  {\includegraphics[width=1.05\columnwidth]{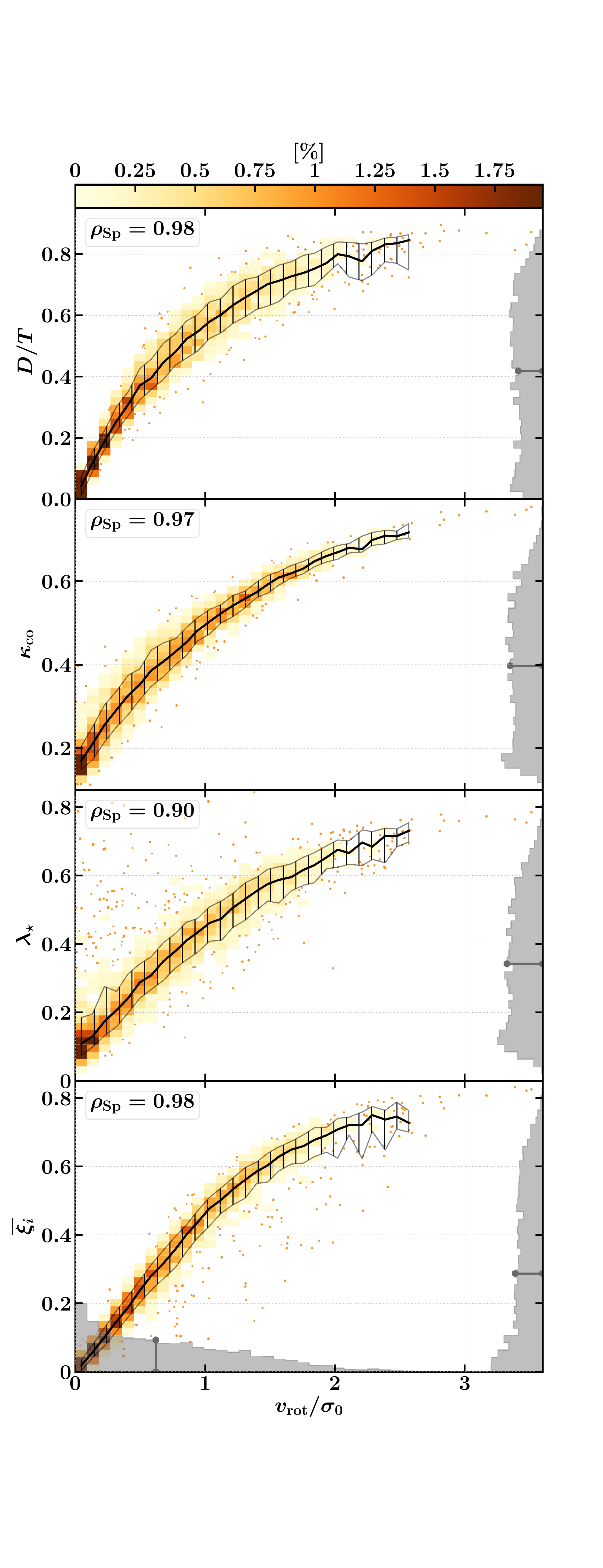}}
\caption{The relationship between $v_{\rm rot}/\sigma_0$ and other kinematic diagnostics commonly used to characterise simulated galaxies, from top to bottom: $D/T$, $\kappa_{\rm co}$, $\lambda_\star$ and $\overline{\xi_i}$. The panels show the 2-dimensional histogram of the $4155$ galaxies in our sample, with the parameter space sampled by 40 cells in each dimension. Cells sampled by at least 3 galaxies are coloured to show their contribution to the distribution; galaxies associated with poorly-sampled cells are drawn individually. Overplotted lines show the binned median and $1\sigma$ ($16^{\rm th}$-$84^{\rm th}$) percentile scatter of the dependent variables. The 1-dimensional distributions in each variable are shown as grey-scale linear histograms, with the medians of these denoted by overlaid signposts. The four dependent variables each correlate strongly with $v_{\rm rot}/\sigma_0$, having Spearman rank-order coefficients, $\rho_{\rm Sp}$, of $0.98$ ($D/T$), $0.97$ ($\kappa_{\rm co}$), $0.90$ ($\lambda_\star$) and $0.98$ ($\overline{\xi_i}$).}
\label{fig:kinem_metrics}
\end{figure}

We show in Fig. \ref{fig:kinem_metrics} the kinematical properties of EAGLE galaxies, as characterised by the diagnostics presented in Section \ref{sec:kin_diag}. From top to bottom, these are: the disc-to-total stellar mass ratio, $D/T$; the kinetic energy in co-rotation, $\kappa_{\rm co}$; the stellar mass-weighted spin parameter, $\lambda_{\star}$; and the median orbital circularity, $\overline{\xi}$. These quantities are shown as a function of the ratio of rotation and dispersion velocities, $v_{\rm rot}/\sigma_0$. The panels show the distribution of the 4155 galaxies of our sample as a 2-dimensional probability distribution function, with 40 cells in each dimension. Only cells sampled by at least 3 galaxies are coloured; galaxies associated with poorly-sampled cells are drawn individually. The overplotted lines show the binned median and $1\sigma$ ($16^{\rm th}$-$84^{\rm th}$ percentile) scatter of the dependent variables. The 1-dimensional distributions in each variable are shown via the grey-scale linear histograms. The median values of $D/T$, $\kappa_{\rm co}$, $\lambda_\star$, $\overline{\xi}$ and $v_{\rm rot}/\sigma_0$, denoted by signposts on the grey-scale histograms, are $0.42$, $0.40$, $0.34$, $0.27$ and $0.62$, respectively.

Reassuringly, there is a strong positive correlation between each of $D/T$, $\kappa_{\rm co}$, $\lambda_\star$ and $\overline{\xi}$, plotted as dependent variables, and $v_{\rm rot}/\sigma_0$. Since the correlations are not linear for all values of $v_{\rm rot}/\sigma_0$, we quantify their strength with the Spearman rank-order coefficient, $\rho_{\rm Sp}$, the values of which are unsurprisingly high: $0.98$ for $D/T$, $0.97$ for $\kappa_{\rm co}$, $0.90$ for $\lambda_\star$, and $0.98$ for $\overline{\xi}$. The scatter at fixed $v_{\rm rot}/\sigma_0$ is greatest for $\lambda_\star$, highlighting the intrinsic uncertainty associated with the recovery of kinematic diagnostics from surface-brightness-limited observations. In contrast, $D/T$, $\kappa_{\rm co}$ and $\overline{\xi_i}$ all scale nearly linearly with $v_{\rm rot}/\sigma_0$ in the regime $v_{\rm rot}/\sigma_0 \lesssim 1$, and $\kappa_{\rm co}$ in particular exhibits relatively little scatter at fixed $v_{\rm rot}/\sigma_0$.

We conclude from this brief examination that the five kinematical diagnostics are broadly consistent and can in general be used interchangeably. Following the suggestion of \citet{Correa_et_al_17} that division of the EAGLE population about a threshold of $\kappa_{\rm co}=0.4$ separates the star-forming and passive galaxy populations (which we show later in Fig. \ref{fig:colour_mag}), we infer that a similar outcome can be achieved by division about a threshold of $v_{\rm rot}/\sigma_0 \simeq 0.7$, which corresponds to $D/T \simeq 0.45$, $\lambda_\star \simeq 0.35$ or  $\overline{\xi_i} \simeq 0.3$. Hereafter, we use $v_{\rm rot}/\sigma_0$ to characterise the internal kinematics of EAGLE galaxies; the main advantages being that it is derived using the same framework with which we compute the velocity dispersion anisotropy, $\delta$, and that it is analogous to observational measurements derived from spectroscopy. 

\begin{figure*}
\centering
\adjustbox{trim={.07\width} {.06\height} {.09\width} {.08\height},clip}%
  {\includegraphics[width=\paperwidth,keepaspectratio]{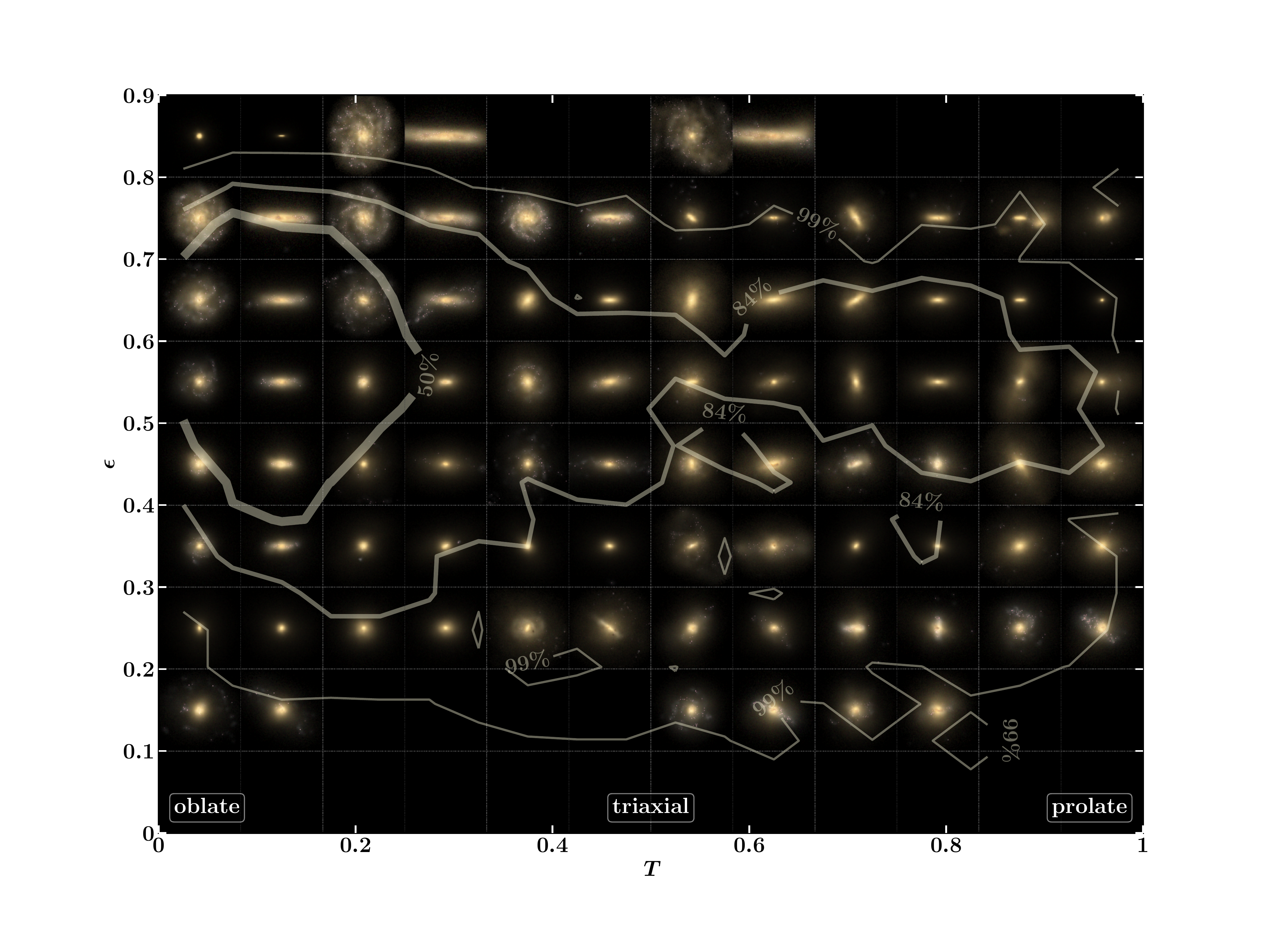}}
 \caption{Two-dimensional histogram of the sample of $4155$ central galaxies in the parameter space defined by the triaxiality, $T$, and flattening, $\epsilon$, shape parameters (see eq. \ref{eq:shapeparameters}). The parameter space is sampled with cells of $\Delta T = 0.05$ and $\Delta \epsilon = 0.04$, and the overlaid contours correspond to the $50^{\rm th}$, $84^{\rm th}$ and $99{\rm th}$ percentiles of the distribution. The background is comprised of pairs of face-on and edge-on images of randomly-selected galaxies, $60 \pkpc$ on a side, drawn from the corresponding region of the parameter space to provide a visual impression of the morphology defined by the corresponding shape parameters. The most common configuration is a flattened, oblate ellipsoid, but galaxies span the majority of the available parameter space.}
 \label{fig:morphology}
\end{figure*}

\section{The morphology and kinematics of EAGLE galaxies}
\label{sec:morphokinem}

We first examine in Section \ref{sec:morphology} the morphology of EAGLE galaxies. In Section \ref{sec:colour_separation} we perform a brief check of how well our chosen diagnostics are able to distinguish between central galaxies in the `blue cloud' and on the `red sequence'. Then in Section \ref{sec:m_k_relationship} we investigate the relationship between the two properties.

\subsection{The morphologies of EAGLE galaxies}
\label{sec:morphology}

Fig. \ref{fig:morphology} shows with contours the 2-dimensional probability distribution of our sample of central galaxies in the space defined by the triaxiality, $T$, and flattening, $\epsilon$, shape parameters (see equation \ref{eq:shapeparameters}). The galaxies are assigned to a grid with 20 cells in each dimension, and contours are drawn for levels corresponding to the \nth{50}, \nth{84} and \nth{99} percentiles of the distribution. The galaxies were then rebinned to a coarse grid of 8 $\times$ 6 cells, and a galaxy within each cell was selected at random. Face-on and edge-on images of these galaxies, created using the techniques described by \citet{Trayford_et_al_15}, were extracted from the EAGLE public database\footnote{http://galaxy-catalogue.dur.ac.uk} \citep{McAlpine_et_al_16} and are shown in the background of the plot to provide a visual impression of the morphology corresponding to particular values of the shape parameters $(\epsilon, T)$. 

\begin{figure*}
\adjustbox{trim={.06\width} {.03\height} {.09\width} {.09\height},clip}%
  {\includegraphics[width=0.8\paperwidth]{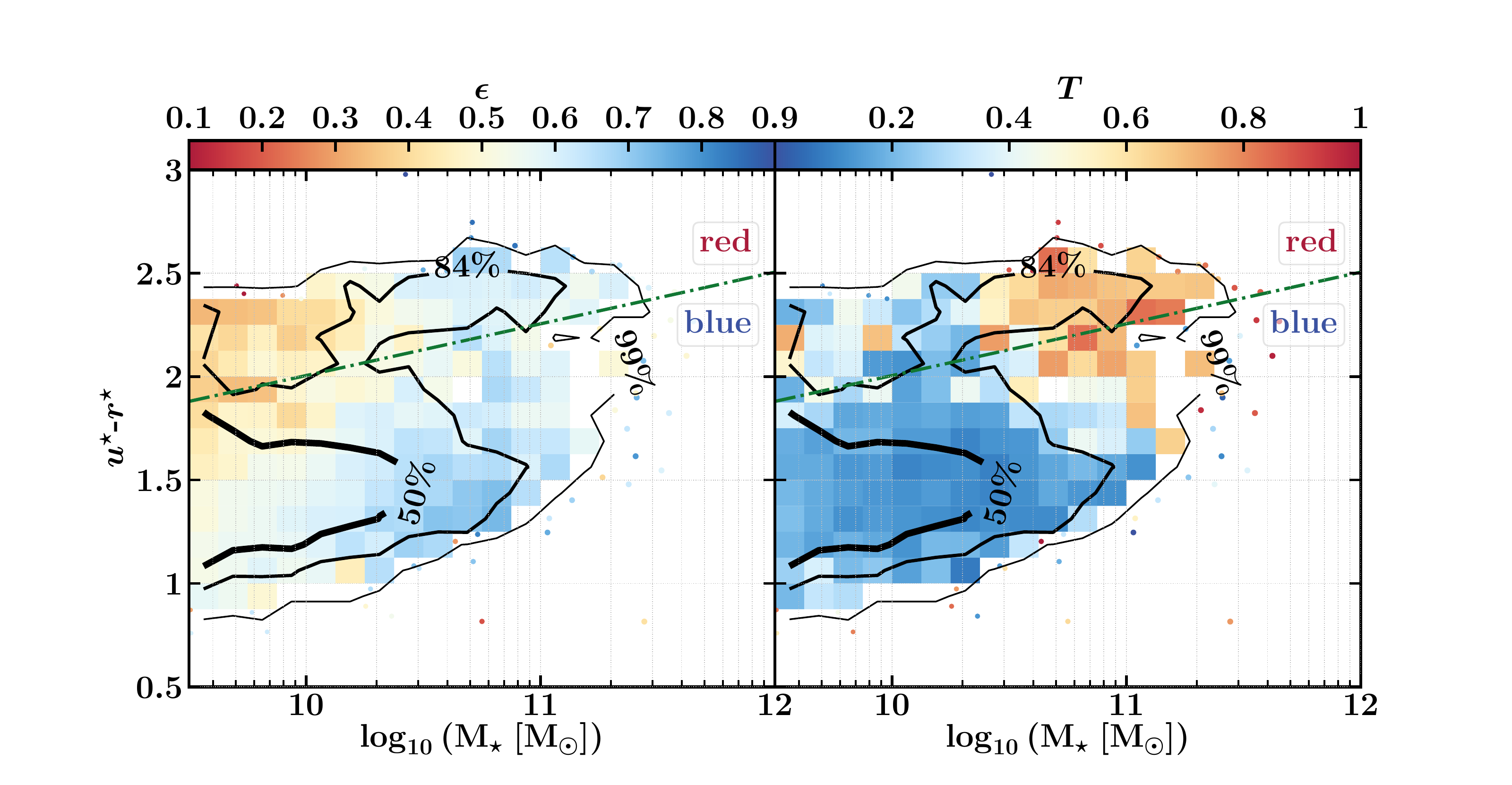}}
\adjustbox{trim={.06\width} {.03\height} {.09\width} {.06\height},clip}%
  {\includegraphics[width=0.8\paperwidth]{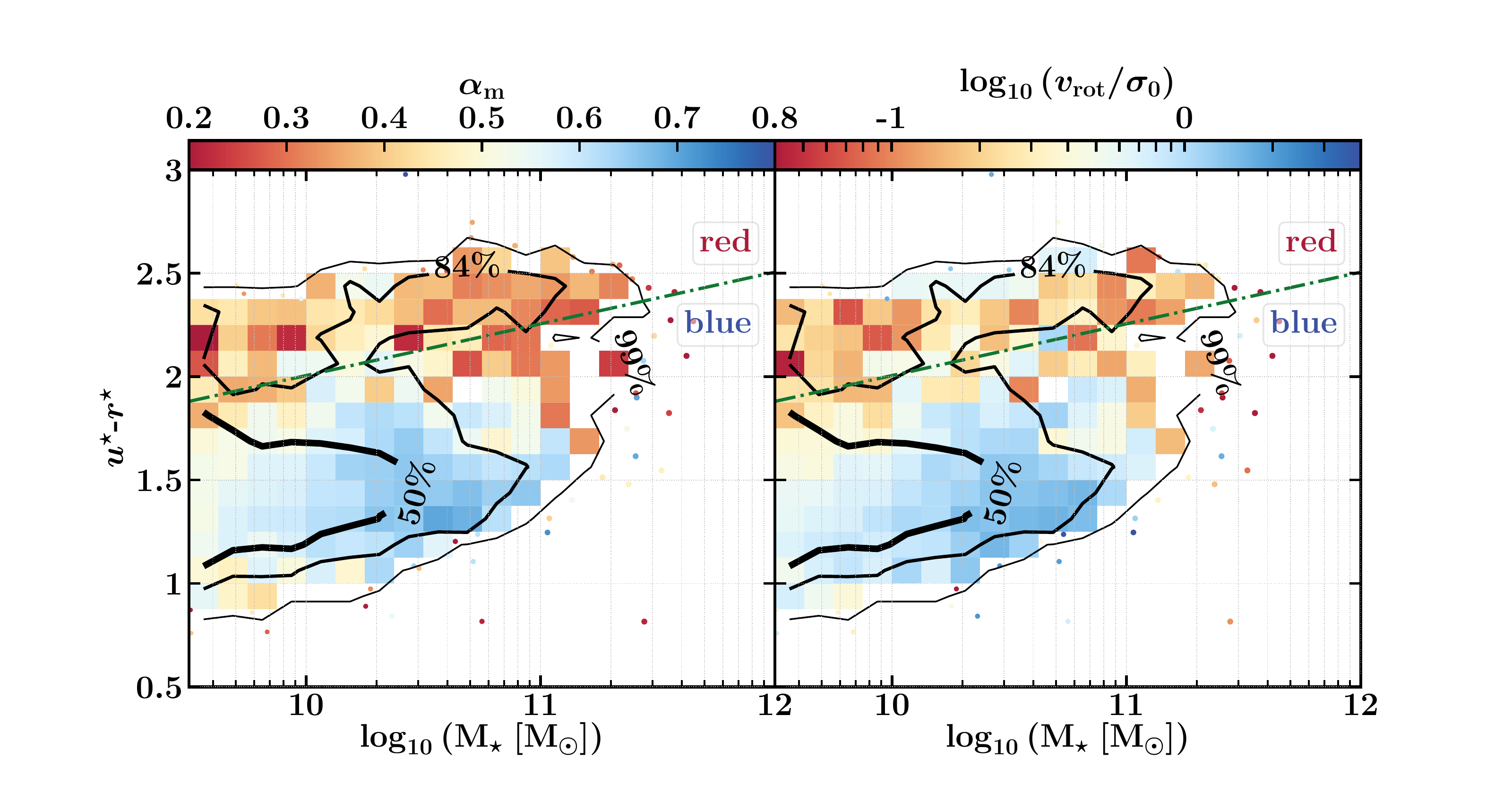}}
\caption{Two-dimensional probability distribution functions of the present-day $(u^\star - r^\star) - M_\star$ relation defined by our sample of $4155$ well-sampled central galaxies. The parameter space is sampled with 20 cells in each dimension, and the overlaid contours correspond to the $50^{\rm th}$, $84^{\rm th}$ and $99{\rm th}$ percentiles of the distribution. In the regime where cells are sampled by fewer than 3 galaxies, galaxies are represented individually by points. The dot-dashed line corresponds to the definition of the `green valley' advocated by \citet{Schawinski_et_al_14}, separating the `red sequence' from the `blue cloud'. Cells and points are coloured by the median flattening, $\epsilon$, of the galaxies in the upper left-hand panel, the median triaxiality, $T$, of galaxies in the upper right-hand panel, by the median of the parameter $\alpha_{\rm m}$ (see main text) in the lower left-hand panel, and by the median rotation-to-dispersion velocity, $v_{\rm rot}/\sigma_0$, in the lower right-hand panel. The colouring shows that the blue cloud is preferentially comprised of flattened, rotationally-supported galaxies with low triaxiality, whilst red sequence galaxies tend to be spheroidal or prolate, and exhibit significant dispersion support. The two populations can be differentiated with similar efficacy using thresholds of $\alpha_{\rm m}\simeq 0.5$ or $v_{\rm rot}/\sigma_0 \simeq 0.7$.}
\label{fig:colour_mag}
\end{figure*}

Oblate systems are found towards the left-hand side of the figure ($T \simeq 0$), and prolate to the right ($T \simeq 1$)\footnote{Ellipsoids with $T \simeq 0.5$ are purely triaxial.}, while highly-flattened discs are found towards the top of the figure and vertically-extended galaxies at the bottom. The contours indicate that the region of the $(\epsilon, T)$ plane most populated by the galaxies satisfying our initial selection criteria is that of flattened, oblate ellipsoids. The sample spans the full range of both parameters, with high-$T$ galaxies tending to be less flattened and hence significantly prolate. The median values of $\epsilon$ and $T$ are $0.46$ and $0.19$, respectively. The plane features two `zones of avoidance', firstly at (high-$\epsilon$, high-$T$) which requires that the entire galaxy assume a bar-like configuration, and secondly at the very lowest flattening values ($\epsilon < 0.1$), which require that galaxies are almost perfectly spherical.

The oblate galaxies exhibit axisymmetry about the minor axis, while prolate systems are characterised by an intermediate axis that is significantly shorter than their major axis, and thus resemble cigars. We note that the face-on and edge-on orientations of the galaxy images were defined relative to the axis of rotation, rather than the structural minor axis; the two axes tend to be near-parallel in relaxed oblate systems but are often mis-aligned in prolate systems, the majority of which rotate about the major rather than minor axis \citep[consistent with the observational findings of][]{Krajnovic_et_al_18}. As such, the images of prolate systems can appear poorly aligned. The images of several prolate systems also show evidence of tidal disturbance and/or merger remnants, suggesting that prolate structure may be induced by interactions with neighboring galaxies. 

Inspection of Fig. \ref{fig:morphology} also highlights a qualitative trend: star-forming galaxies, which are identifiable via blue, and typically extended, components in the images, are found preferentially in the (high-$\epsilon$, low-$T$) regime, characteristic of discs. Conversely, red quiescent galaxies are preferentially located the low-$\epsilon$ regime. However, we note that the images of galaxies in the prolate regime exhibit blue, often asymmetric, structures, indicating that the interactions with neighbouring galaxies that induce prolate structure also induce star formation. \citet{Trayford_et_al_16} show that such interactions can enable red galaxies to temporarily ``rejuvenate'', and move from the red sequence back to the blue cloud \citep[see also e.g.][]{Robertson_et_al_06}. In the following section, we explore the consequences of this complexity in the ($\epsilon$, $T$) plane in more detail.

\subsection{Correspondence with the colour-mass relation}
\label{sec:colour_separation}

We now turn to a quantitative examination of the relationship between the morphology and kinematics of galaxies on the one hand, and their location in the colour-mass plane on the other hand. The panels of Fig. \ref{fig:colour_mag} show with contours the 2-dimensional histogram of the simulated galaxies in the $(u^\star-r^\star)$ colour - stellar mass plane, where the superscript $^\star$ denotes intrinsic colours, i.e. rest-frame and dust-free. In all panels, galaxies are binned onto a grid with 20 cells in each dimension. As per the images shown in Fig. \ref{fig:morphology}, broadband magnitudes were retrieved from the EAGLE public database, having been computed with the techniques described by \citet{Trayford_et_al_15}, who showed that the $(g-r)$ colours of the  EAGLE galaxy population are consistent with the dust-corrected colours of observed galaxies. 

The dashed line overlaid on each panel corresponds to the definition of the `green valley' proposed by \citet{Schawinski_et_al_14}, $(u^\star-r^\star) = 0.25\log_{10}(M_\star/{\rm M}_\odot)-0.495$, which separates the blue cloud of star-forming galaxies from the red sequence of passive galaxies. \citet{Trayford_et_al_15,Trayford_et_al_17} show that EAGLE's galaxy population naturally divides into these two populations, and we see here that, despite the omission of satellite galaxies from our sample, which comprise a significant fraction of the low-mass regime of the present-day red sequence, colour bimodality is still clearly visible in the contours.

In the upper left-hand panel of Fig. \ref{fig:colour_mag}, well-sampled cells are coloured by the median value of the flattening parameter, $\epsilon$, of the galaxies within the cell. This plot is therefore analogous to the colour-mass diagram shown by \citet[][their Fig. 3]{Correa_et_al_17}, in which the galaxies are coloured by $\kappa_{\rm co}$. As noted in Section \ref{sec:compare_kinematics}, those authors show that the blue cloud and red sequence can be reasonably well separated by a simple threshold in $\kappa_{\rm co}$. As might also be inferred from inspection of Fig. \ref{fig:morphology}, we find here that a simple threshold in $\epsilon$ does not enable such a clean separation; whilst the blue cloud is dominated by flattened galaxies $\epsilon \gtrsim 0.5$, only the low-mass end of the red sequence is dominated by spheroidal galaxies. Inspection of the upper right-hand panel of Fig. \ref{fig:colour_mag}, in which the cells are coloured by the median value of the triaxiality parameter, $T$, shows that the flattened galaxies populating the high-mass end of the red sequence are prolate ($T \simeq 1$) rather than disc-dominated systems. An increasing prolate fraction with increasing stellar mass was also reported by \citet{Li_et_al_18} based on an analysis of the morphology of galaxies in the Illustris simulations, and recent observations with the MUSE integral field spectrograph of massive galaxies corroborate this prediction \citep[][]{Krajnovic_et_al_18}. Conversely, we find that the blue cloud is overwhelmingly dominated by flattened systems with very low values of the triaxiality parameter, i.e. disky galaxies. 

Since neither $\epsilon$ nor $T$ alone affords a simple means of separating the blue cloud from the red sequence, we construct a new morphological diagnostic that combines both shape parameters, $\alpha_{\rm m} = (\epsilon^2 + 1-T)/2$. By construction, this diagnostic separates spheres and prolate spheroids, characteristic of the morphology of early-type galaxies, from the oblate spheroids that characterise the morphology of late-type galaxies. Cells are coloured by this quantity in the lower left-hand panel of Fig. \ref{fig:colour_mag}, showing that a simple threshold of $\alpha_{\rm m} \simeq 0.5$, does a reasonable job of distinguishing galaxies of the blue cloud from those of the red sequence.

In the lower right-hand panel of Fig. \ref{fig:colour_mag}, cells are coloured by the median $v_{\rm rot}/\sigma_0$ of the galaxies in each pixel. Visual inspection shows that the blue cloud is dominated by rotationally-supported galaxies, whilst the galaxies comprising the red sequence are generally dispersion supported. As might be expected when considering the correspondence between $\kappa_{\rm co}$ and $v_{\rm rot}/\sigma_0$ discussed in Section \ref{sec:compare_kinematics}, a simple threshold in the latter (e.g. $v_{\rm rot}/\sigma_0 \simeq 0.7$) therefore differentiates the blue cloud from the red sequence with a similar efficacy to the $\kappa_{\rm co}=0.4$ threshold advocated by \citet{Correa_et_al_17}. Comparison of the lower two panels of Fig. \ref{fig:colour_mag} thus shows that the two populations can be differentiated with similar efficacy using morphological or kinematical diagnotics, but only for the former if one is able to measure both the flattening and triaxiality of the galaxies. In general this is not the case for observational studies, making the spectroscopically-accessible $v_{\rm rot}/\sigma_0$ diagnostic a particularly effective means of differentiation. For analyses of simulations, the $\kappa_{\rm co}$ diagnostic is attractive, since it requires the calculation of a single quantity that is simple to interpret and which takes values in the range [0,1].

\subsection{The relationship between morphology and kinematics}
\label{sec:m_k_relationship}

\begin{figure}
\centering
\adjustbox{trim={.03\width} {.08\height} {.07\width} {.09\height},clip}%
  {\includegraphics[width=1.11\columnwidth,keepaspectratio]{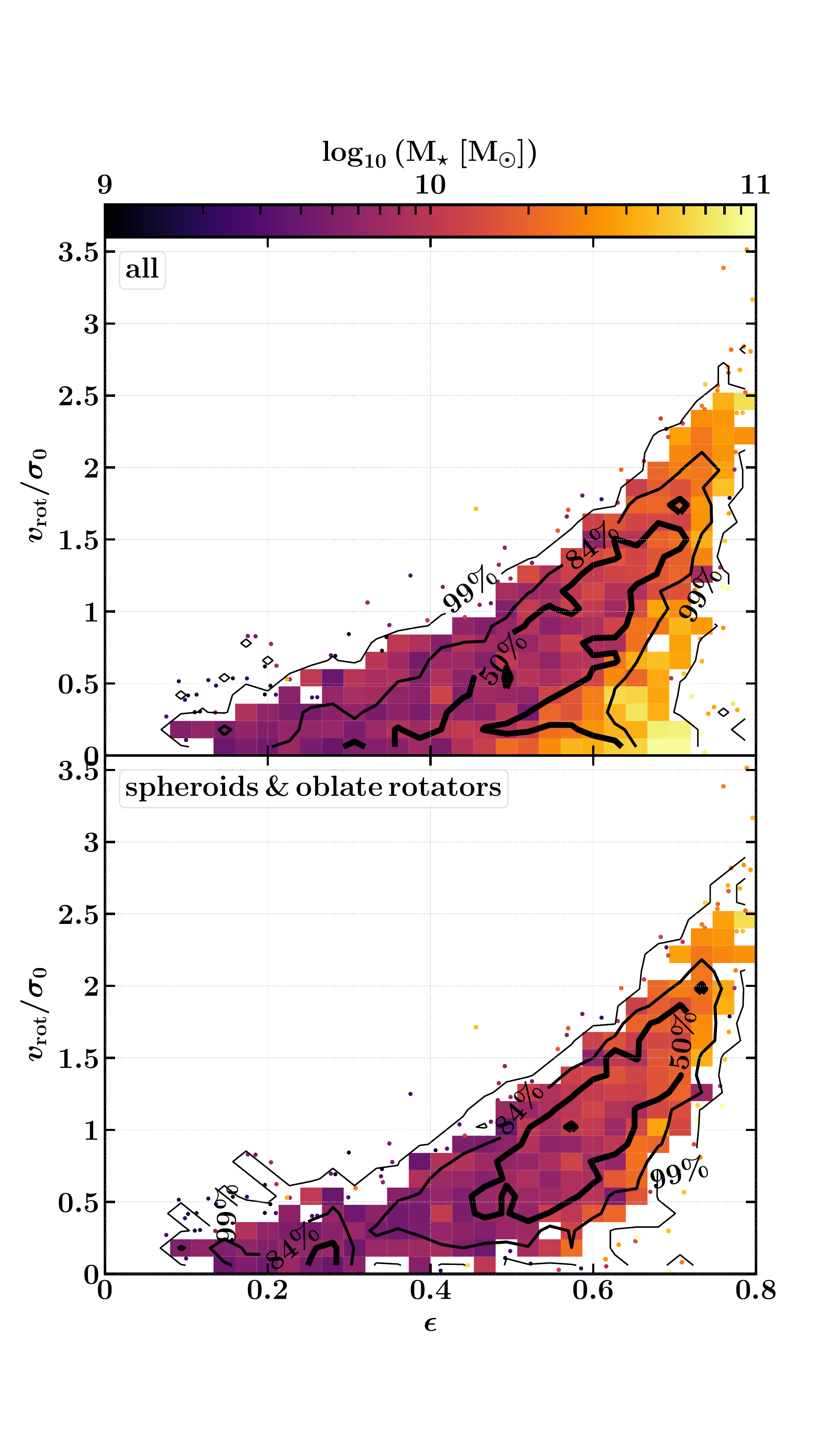}}
\caption{Two-dimensional probability distribution function in the $v_{\rm rot}/\sigma_0$ and flattening, $\epsilon$, plane shown as contours corresponding to the $50^{\rm th}$, $84^{\rm th}$ and $99{\rm th}$ percentiles of the distribution, drawn on a 30 $\times$ 30 grid. Well-sampled cells are colour-coded by their median stellar masses, while galaxies associated with cells sampled by fewer than 3 galaxies are drawn and coloured individually. The upper panel shows the full sample of 4155 galaxies, the lower panel shows the subset of 2703 galaxies that are spheroidal ($\epsilon < 0.3$), or oblate ($T < 1/3$) and have their angular momentum axis aligned with their structural minor axis to within 10 degrees. Excision of the prolate and mis-aligned systems primarily eliminates a population of dispersion-supported ($v_{\rm rot}/\sigma_0 \simeq 0$) galaxies with diverse morphologies. The remaining sample exhibits a strong correlation between the morphological and kinematic diagnostics, but there is significant scatter in $\epsilon$ at fixed $v_{\rm rot}/\sigma_0$.}
\label{fig:m_k_samples}
\end{figure}

We now turn to the correspondence between the morphology and kinematics of EAGLE galaxies. In Fig. \ref{fig:m_k_samples}, we show how galaxies populate the $v_{\rm rot}/\sigma_0 - \epsilon$ plane with contours corresponding to the $50^{\rm th}$, $84^{\rm th}$ and $99{\rm th}$ percentiles of the distribution. The background colouring denotes the characteristic stellar mass of galaxies at each location in $v_{\rm rot}/\sigma_0 - \epsilon$ space. \citet{van_de_Sande_et_al_18a} recently demonstrated that EAGLE galaxies selected to mimic those targeted by the SAMI survey populate this parameter space in a similar fashion to the observed sample. The upper panel of Fig. \ref{fig:m_k_samples} shows the distribution of all $4155$ central galaxies comprising our overall sample, whilst the lower panel shows the distribution for the sub-set of $2703$ galaxies that are spheroidal ($\epsilon < 0.3$), or are oblate ($T < 1/3$) and have their angular momentum axis aligned with their structural minor axis to within 10 degrees. As might be inferred from Fig. \ref{fig:colour_mag}, massive galaxies (i.e. $M_\star \gtrsim 10^{10.5}\Msun$) tend to populate the high-$\epsilon$ regime, but exhibit a diverse range of $v_{\rm rot}/\sigma_0$ values since they can be rotating discs or prolate spheroids with significant dispersion support. The excision of prolate galaxies, and a small number of systems whose morphology and kinematics have been influenced significantly by encounters with neighbours or recently-merged satellites, therefore preferentially eliminates a population of dispersion-supported ($v_{\rm rot}/\sigma_0 \simeq 0$) galaxies with diverse morphologies. 

The remaining sample exhibits a strong correlation between the morphological and kinematic diagnostics (Spearman rank-order coefficient of $\rho_{\rm Sp} = 0.72$), but with significant scatter in $\epsilon$ at fixed $v_{\rm rot}/\sigma_0$. It is possible to identify galaxies with $v_{\rm rot}/\sigma_0 \simeq 1$ and flattening parameters as diverse as $\epsilon \simeq 0.3-0.8$. Similarly, flattened galaxies with $\epsilon \simeq 0.7$ can exhibit rotation-to-dispersion velocity ratios between $v_{\rm rot}/\sigma_0 \simeq 0.5$ and $v_{\rm rot}/\sigma_0 \simeq 3.5$. It is therefore clear that morphological and kinematical diagnostics are not trivially interchangeable, indicating that the morphology of a galaxy is significantly influenced not only by $v_{\rm rot}/\sigma_0$, but also by at least one other property. In this respect the simulations are in qualitative agreement with the findings of surveys conducted with panoramic integral field spectrographs \citep[see e.g. the review by][]{Cappellari_16}.

\subsubsection{The influence of velocity dispersion anisotropy}
\label{sec:anisotropy}

\begin{figure*}
\centering
\adjustbox{trim={.01\width} {.01\height} {.05\width} {.08\height},clip}%
  {\includegraphics[width=0.5\textwidth,keepaspectratio]{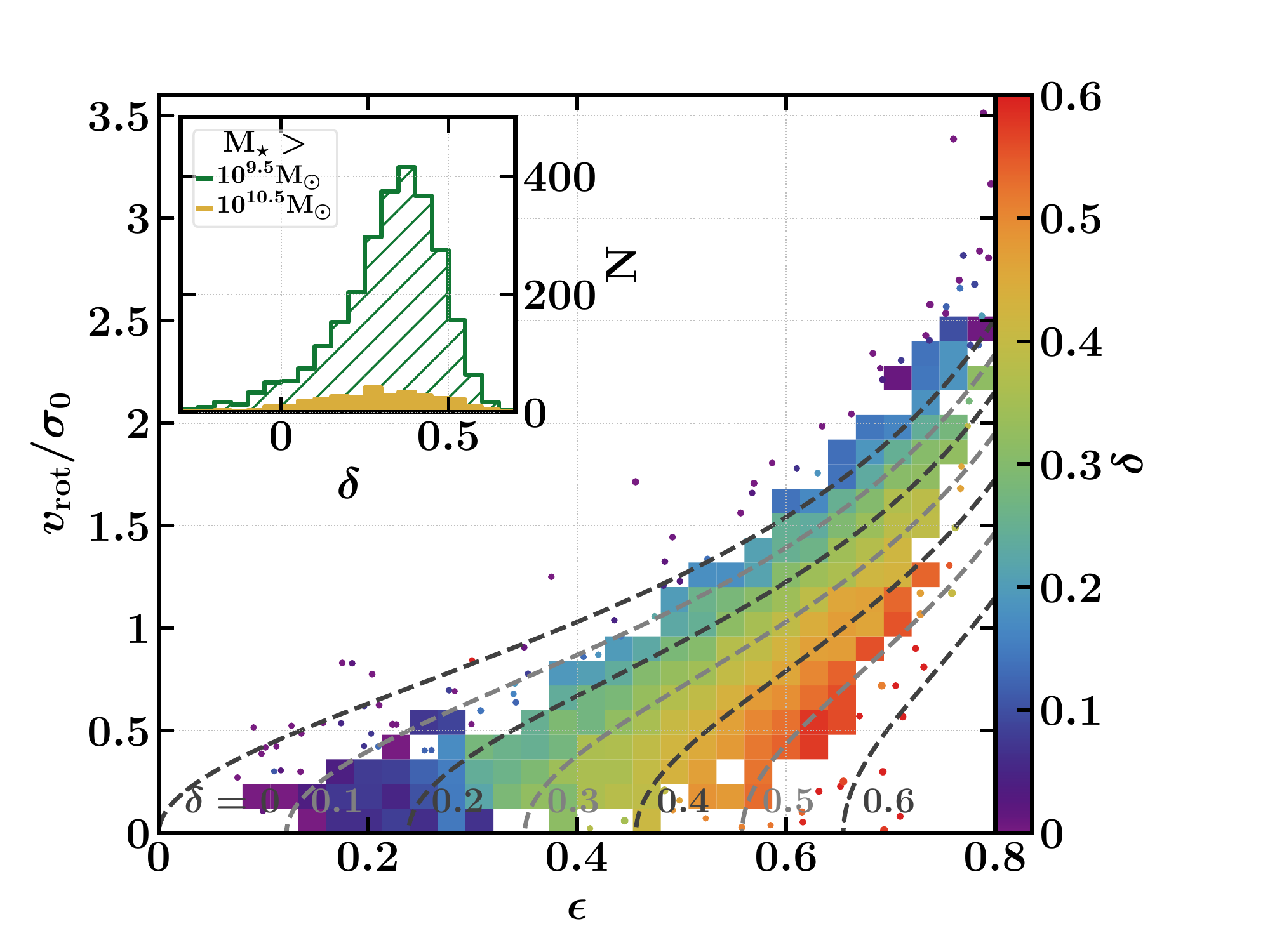}}
\adjustbox{trim={.00\width} {.01\height} {.05\width} {.08\height},clip}%
  {\includegraphics[width=0.5\textwidth,keepaspectratio]{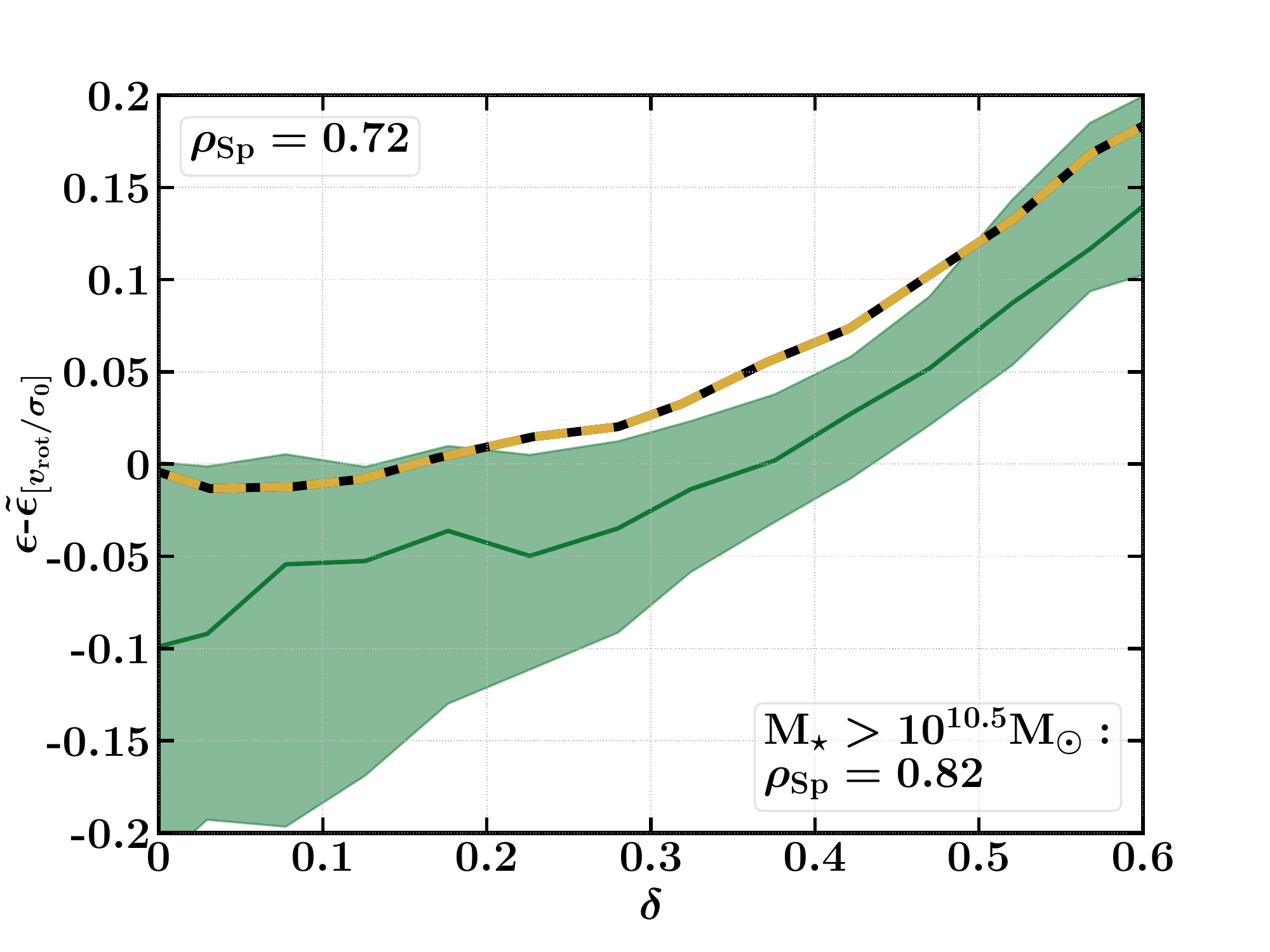}} 
 \caption{\textit{Left:} The panel shows the same sample of galaxies shown in the lower panel of Fig. \ref{fig:m_k_samples}, but here the colour coding of each cell denotes the median dispersion anisotropy $\delta$ (eq. \ref{eq:anisotropy_param}) of its associated  galaxies. The alternately black and grey dashed curves represent the $v_{\rm rot}/\sigma_0$ - $\epsilon$ relation expected for $\delta = 0.0-0.6$ in increments of $0.1$, from application of the tensor virial theorem. The simulations reproduce the analytical predictions in both qualitative and quantitative senses, with increased flattening at fixed $v_{\rm rot}/\sigma_0$ clearly associated with increased anisotropy. The inset panel shows the histogram of anisotropy values realised by this sample (green hatching) and also separately for galaxies with $\rm{M_{\star}} > 10^{10.5} \Msun$ (yellow). \textit{Right:} the deviation of a galaxy's flattening, $\epsilon$, from the median flattening for galaxies of similar $v_{\rm rot}/\sigma_0$, $\tilde{\epsilon}_{[v_{\rm rot}/\sigma_0]}$, as a function of $\delta$. The solid line and shaded region denote the median and $1\sigma$ ($16^{\rm th}$-$84^{\rm th}$ percentile) scatter about it, respectively, in bins of $\Delta \delta = 0.05$. The correlation of these quantities has a Spearman rank-order coefficient of $\rho_{\rm Sp} = 0.72$. Similarly, the yellow dashed curve represents the median relationship for the high-mass end sub-population, showing a correlation with a Spearman rank-order coefficient of $\rho_{\rm Sp} = 0.82$.}
\label{fig:m_k_anisotropy}
\end{figure*}

The morphology and kinematics of collisionless systems are linked via the tensor virial theorem. Its application to oblate, axisymmetric spheroids rotating about their short axis, modelled as collisionless gravitating systems, is discussed in detail by \citet{Binney_78} and \citet{Binney_and_Tremaine_87}. They show that such bodies trace distinct paths in the $v_{\rm rot}/\sigma_0$ - $\epsilon$ plane, for fixed values of the velocity dispersion anisotropy, $\delta$ (see eq. \ref{eq:anisotropy_param}), offering a potential explanation for the morphological diversity of galaxies at fixed $v_{\rm rot}/\sigma_0$.

In the left-hand panel of Fig. \ref{fig:m_k_anisotropy}, we plot once again the $v_{\rm rot}/\sigma_0 - \epsilon$ distribution of the sub-sample of $2703$ spheroidal and well-aligned oblate galaxies. Here, the colour coding denotes the median velocity dispersion anisotropy of galaxies associated with each cell that was shown in the bottom panel of Fig. \ref{fig:m_k_samples}. The overlaid dashed curves represent the $v_{\rm rot}/\sigma_0 - \epsilon$ relation expected from application of the tensor virial theorem to collisionless gravitating systems with $\delta = 0.1 - 0.6$, in increments of $0.1$ (for which a derivation is provided in Appendix \ref{sec:virial}). We remind the reader that, as shown in Fig. \ref{fig:m_k_samples}, the high-$\epsilon$ regime is dominated by more-massive galaxies ($M_\star \gtrsim 10^{10.5}\Msun$), whilst the majority of the plane is sampled by galaxies with mass closer to our selection limit of $M_\star = 10^{9.5}\Msun$. 

The main plot demonstrates that the analytic predictions are a good representation of the behaviour of the simulated galaxies. At fixed $v_{\rm rot}/\sigma_0$, more anisotropic galaxies are clearly associated with a more flattened morphology; taking galaxies with $v_{\rm rot}/\sigma_0 \simeq 1$ as an example, those with $\epsilon \simeq 0.45$ exhibit a typical anisotropy of $\delta \simeq 0.2$, whilst the most-flattened examples, with $\epsilon \simeq 0.7$, exhibit $\delta \simeq 0.5$. The inset panel shows the histogram of anisotropy values realised by the sub-sample of $2703$ (green hatching), and also those of the $329$ galaxies from this sub-sample with $M_\star \gtrsim 10^{10.5}\Msun$ (yellow). For the main sub-sample, the distribution is broadly symmetric about a median of $0.34$, albeit with a more extended tail to low (even negative) values. A small but significant fraction of galaxies in the sample ($\simeq$ 5 percent), exhibit $\delta > 0.5$. The subset of high-mass galaxies spans a similar range in $\delta$ but exhibits a lower median value of $0.29$. 

To highlight the influence of $\delta$ on morphology more clearly, we compute $\tilde{\epsilon}_{[v/\sigma]}$, the median flattening parameter of galaxies in bins of fixed $v_{\rm rot}/\sigma_0$, and plot in the right-hand panel of Fig. \ref{fig:m_k_anisotropy}, as a function of $\delta$, the deviation of each galaxy's flattening parameter from this median, $\epsilon - \tilde{\epsilon}_{[v/\sigma]}$. The solid line and shaded region denote the median and $1\sigma$ ($16^{\rm th}-84^{\rm th}$ percentile) scatter of this deviation in bins of $\Delta \delta = 0.05$. The two quantities are strongly correlated, with a Spearman rank-order coefficient of $\rho_{\rm Sp} = 0.72$. The yellow dashed curve represents the trend of the sub-set of high-mass galaxies; since these galaxies are largely confined to high-$\epsilon$ values, we cannot recompute $\tilde{\epsilon}_{[v/\sigma]}$ from this sub-set, and use that of the main sample. As such the median $\epsilon - \tilde{\epsilon}_{[v/\sigma]}$ at fixed $\delta$ for this subset is necessarily elevated. The morphologies of the high-mass subset are even more strongly correlated with the anisotropy, with a Spearman rank-order coefficient of $\rho_{\rm Sp} = 0.82$. The physical interpretation one may therefore draw is that the flattening of EAGLE galaxies, particularly those with low and intermediate levels of rotation support (i.e. $v_{\rm rot}/\sigma_0 < 1$), can be influenced significantly by the anisotropy of the stellar velocity dispersion, with some galaxies exhibiting anisotropy values as high as $\delta \simeq 0.5$. 

\section{The origin of velocity anisotropy }
\label{sec:origin_anisotropy}

\begin{figure}
\centering
\adjustbox{trim={.03\width} {.08\height} {.07\width} {.09\height},clip}%
  {\includegraphics[width=1.11\columnwidth,keepaspectratio]{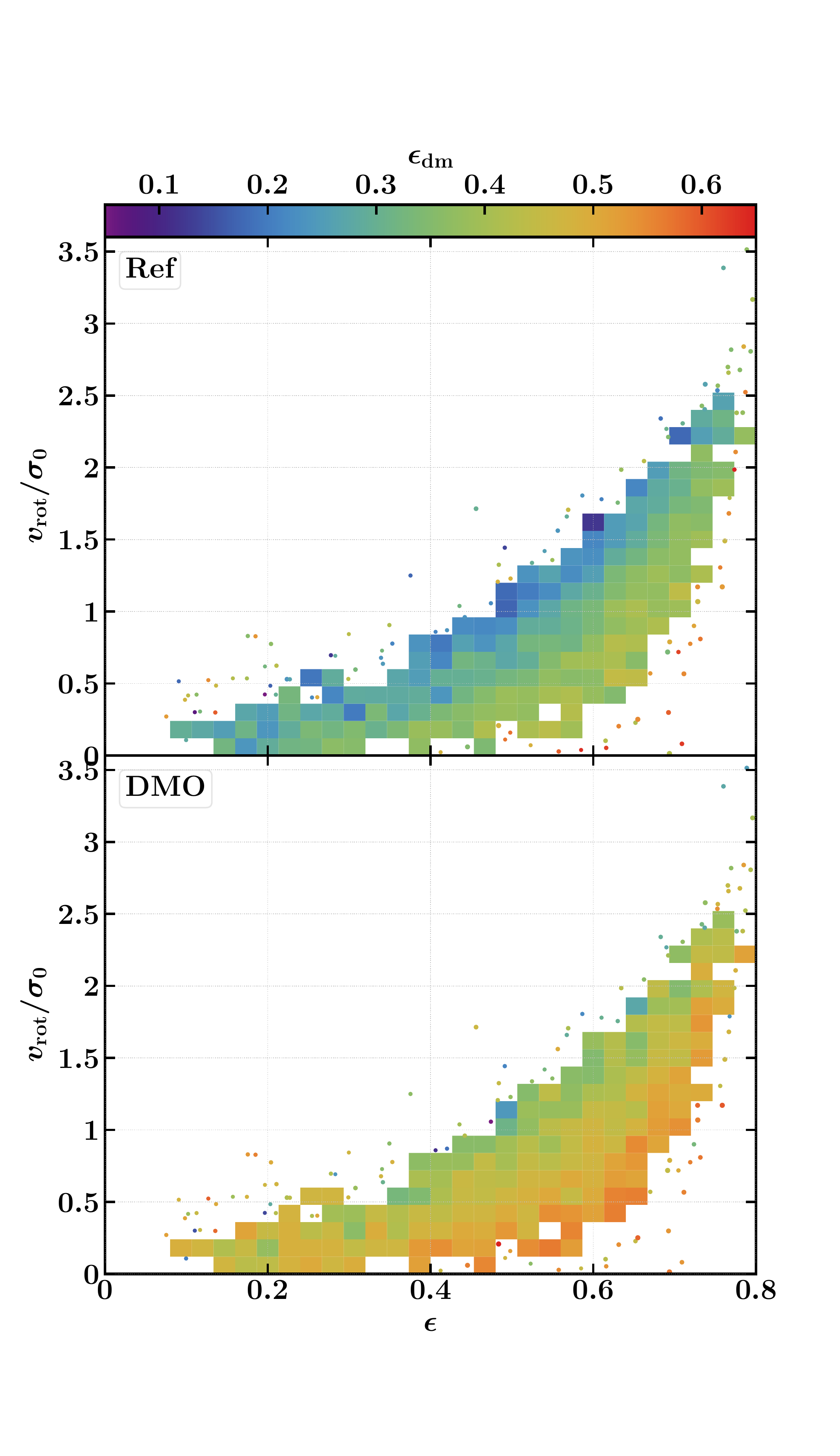}}
\caption{The same sample of galaxies shown in the lower panel of Fig. \ref{fig:m_k_samples}, but here the cells are colour-coded by the flattening of the inner ($<30\pkpc$) dark matter halo, $\epsilon_{\rm dm}$. In the upper panel, this quantity is equated to the flattening of the dark matter halo in the Reference simulation, $\epsilon_{\rm dm}^{\rm Ref}$, whilst in the bottom panel it is equated to the flattening of the corresponding halo in dark matter-only simulation, $\epsilon_{\rm dm}^{\rm DMO}$. Irrespective of which measure is used, the most flattened galaxies at fixed $v_{\rm rot}/\sigma_0$ are preferentially hosted by dark matter haloes whose inner regions are more flattened.}
\label{fig:m_k_eps_for_dm}
\end{figure}

The simulations enable us to examine the origin of the velocity anisotropy that, as discussed in the previous section, can have a significant influence on galaxy morphology. Since the equilibrium orbits of stellar particles are strongly influenced by the structure of the gravitational potential, we focus on the relationship between the velocity anisotropy of galaxies and the morphology of their dark matter haloes, since the latter is a proxy for the structure of the potential.

In analogy with the morphology of galaxies, we quantify the halo morphology via the flattening parameter, $\epsilon_{\rm dm}$, in this case applying the iterative reduced tensor to the distribution of dark matter particles. Since we are concerned with the structure of the potential in the same region for which we have `tracers' of the potential (i.e. stellar particles), we begin iterating the tensor on the set of dark matter particles located within the same $r = 30\,{\pkpc}$ spherical aperture, centred on the galaxy's potential minimum, that is applied to the stellar particles when computing the flattening\footnote{We find that correlations between galaxy morphology or velocity anisotropy with the `global' halo morphology, i.e. considering all dark matter particles bound to the main subhalo, are weak. This is perhaps unsurprising, since the galaxy is most directly influenced by the inner halo \citep{Zavala_et_al_16}, and it is well established that the morphology and kinematics of central galaxies are not strongly correlated with those of their host haloes \citep[e.g.][]{Sales_et_al_12}.}, $\epsilon$. We compute the halo flattening for galaxies in the Ref-L100N1504 simulation, denoting this quantity as $\epsilon_{\rm dm}^{\rm Ref}$, and also for their counterparts identified in a simulation of the same volume, at the same resolution, but considering only collisionless gravitational dynamics (DMONLY-L100N1504). This latter quantity, which we denote as $\epsilon_{\rm dm}^{\rm DMO}$, is instructive because it describes the \textit{intrinsic} shape of the halo that emerges in the absence of the dissipative physics of galaxy formation, and thus enables us to distinguish between cause and effect. The haloes are paired between the Ref and DMONLY simulations using the bijective particle matching algorithm described by \citet{Schaller_et_al_15a}, which successfully pairs $2678$ of the $2703$ haloes that host spheroidal and well-aligned oblate galaxies ($99.1$ percent; see Section \ref{sec:m_k_relationship} for the definition of the sample). 

The panels of Fig. \ref{fig:m_k_eps_for_dm} show the distribution of the matched galaxies in the $v_{\rm rot}/\sigma_0$ - $\epsilon$ plane. Here, the cells and points are coloured by the median value of $\epsilon_{\rm dm}^{\rm Ref}$ and $\epsilon_{\rm dm}^{\rm DMO}$ in the upper and lower panels, respectively. Both panels show a clear trend such that, in the regime of intermediate rotational support, the flattening of the galaxy correlates significantly with the flattening of the central regions of its parent halo, irrespective of whether $\epsilon_{\rm dm}$ is measured in the Ref or DMONLY simulation.

\begin{figure}
\centering
\adjustbox{trim={.0\width} {.07\height} {.07\width} {.09\height},clip}%
  {\includegraphics[width=1.075\columnwidth,keepaspectratio]{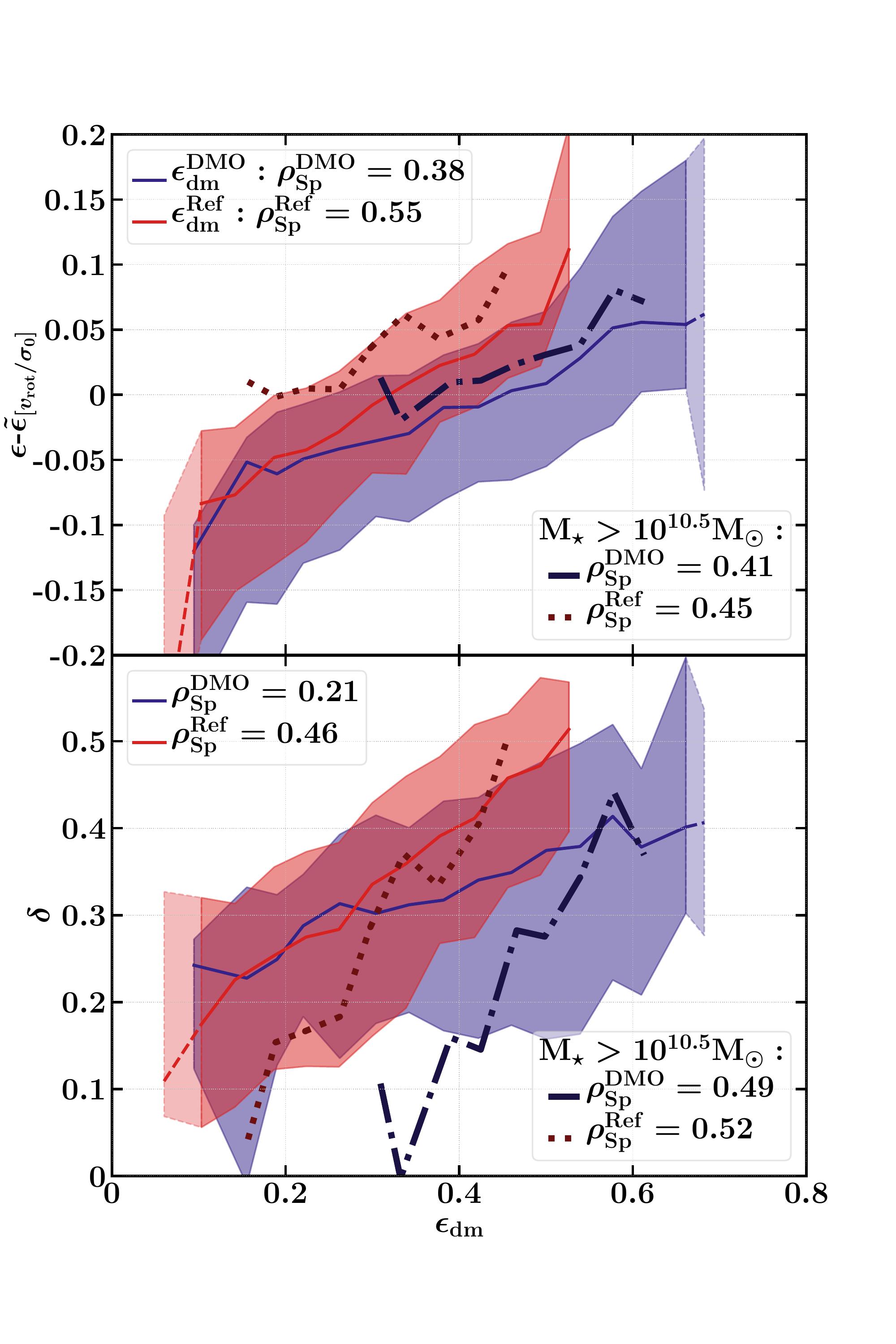}}
\caption{The deviation of a galaxy's flattening, $\epsilon$ from the median flattening for galaxies of similar $v_{\rm rot}/\sigma_0$ ($\epsilon - \tilde{\epsilon}_{[v_{\rm rot}/\sigma_0]}$, \textit{upper panel}), and the anisotropy of the stellar velocity dispersion ($\delta$, \textit{lower panel}), as a function of the flattening of the inner dark matter halo, $\epsilon_{\rm dm}$. The solid lines and shaded regions denote the median relations and the $1\sigma$ ($16^{\rm th}-84^{\rm th}$ percentile) scatter about them, respectively. The red curves adopt $\epsilon_{\rm dm}^{\rm Ref}$ as the halo flattening diagnostic, the blue curves adopt $\epsilon_{\rm dm}^{\rm DMO}$. Medians are drawn with dashed lines in bins sampled by fewer than 5 galaxies. Dotted and dot-dashed thick lines represent the corresponding median relations for the high-mass ($M_\star > 10^{10.5}\Msun$) sub-population for which only bins sampled by more than 5 galaxies are drawn.}
\label{fig:eps_dm_correlations}
\end{figure}

The influence of the morphology of the halo on that of the galaxy is shown more clearly in the upper panel of Fig. \ref{fig:eps_dm_correlations} where, in analogy to the right hand panel of Fig. \ref{fig:m_k_anisotropy}, we show the deviation of a galaxy's flattening parameter from the median flattening of galaxies with similar kinematics, $\epsilon - \tilde{\epsilon}_{[v/\sigma]}$, here as a function of the inner halo flattening. The red curve adopts $\epsilon_{\rm dm}^{\rm Ref}$ as the halo flattening diagnostic, and should be compared to the upper panel of Fig. \ref{fig:m_k_eps_for_dm}, whilst the blue curve adopts $\epsilon_{\rm dm}^{\rm DMO}$ and shows the correlation present in the lower panel of Fig. \ref{fig:m_k_eps_for_dm}. As per the yellow dashed curve of Fig. \ref{fig:m_k_anisotropy}b, dotted and dot-dashed thick lines here denote the median relations constructed using the sub-set of galaxies with $M_\star > 10^{10.5}\Msun$.

The formation of stars following the dissipative collapse of gas drives dark matter haloes towards a more spherical and axisymmetric morphology  \citep[e.g. ][]{Katz_and_Gunn_91,Dubinski_94,Evrard_Summers_and_Davis_94,Springel_et_al_04,Kazantzidis_et_al_04,Bryan_et_al_12,Bryan_et_al_13}, such that in general $\epsilon_{\rm dm}^{\rm DMO} > \epsilon_{\rm dm}^{\rm Ref}$ (see Fig. \ref{fig:m_k_eps_for_dm}). The two halo flattening diagnostics are strongly correlated ($\rho_{\rm Sp} > 0.5$) but the fractional deviation from the 1:1 relation correlates, unsurprisingly, with the halo's stellar mass fraction within $30 \kpc$. The morphological transformation of the halo by dissipative physics therefore acts to compress the dynamic range in $\epsilon_{\rm dm}$, steepening the gradient of the  $(\epsilon - \tilde{\epsilon}_{[v/\sigma]}) - \epsilon_{\rm dm}$ and $\delta - \epsilon_{\rm dm}$ relations. However, the Spearman rank correlation coefficients of these relationships are significantly higher when considering $\epsilon_{\rm dm}^{\rm Ref}$; for the former we recover $\rho_{\rm Sp}=(0.38,0.55)$ for ($\epsilon_{\rm dm}^{\rm DMO}$,$\epsilon_{\rm dm}^{\rm Ref}$) respectively, whilst for the latter relationship we recover $\rho_{\rm Sp}=(0.21,0.46)$. The compression of the dynamic range therefore does not preserve the rank ordering in $\epsilon_{\rm dm}$, and indicates that this property is, perhaps unsurprisingly, not the sole influence on galaxy morphology at fixed $v/\sigma_0$. Nonetheless, the panel shows that, irrespective of which halo flattening diagnostic is considered, there is a clear positive correlation between the morphology of the galaxy (at fixed $v_{\rm rot}/\sigma_0$) and that of its host halo. The persistence of the correlation when considering $\epsilon_{\rm dm}^{\rm DMO}$ demonstrates that it is intrinsic, and does not emerge as a \textit{response} to the formation of a flattened galaxy at the halo centre. This engenders confidence that there is a causal connection between a galaxy's morphology and that of its host inner dark matter halo, which agrees with the findings of \citet{Zavala_et_al_16}. 

Having seen that the deviation of a galaxy's flattening from the median flattening at fixed $v_{\rm rot}/\sigma_0$, $\epsilon - \tilde{\epsilon}_{[v_{\rm rot}/\sigma_0]}$, correlates with both the anisotropy of the stellar velocity dispersion, $\delta$ (see Fig. \ref{fig:m_k_anisotropy}), and the morphology of the dark matter halo, $\epsilon_{\rm dm}$ (see Fig. \ref{fig:m_k_eps_for_dm}), we check the correlation between $\delta$ and $\epsilon_{\rm dm}$, shown in the bottom panel of Fig. \ref{fig:eps_dm_correlations}. Again, dotted and dot-dashed thick lines here correspond to the sub-set of galaxies with $M_\star > 10^{10.5}\Msun$. There is a clear positive correlation between $\delta$ and $\epsilon_{\rm dm}$, which again persists when one considers $\epsilon_{\rm dm}^{\rm DMO}$ rather than $\epsilon_{\rm dm}^{\rm Ref}$, indicative of an intrinsic rather than an induced correlation. We note that in the specific case of high-stellar mass galaxies, haloes whose central regions are relatively unflattened induce significantly less anisotropy than is the case for the broader galaxy population; this is to be expected since high-stellar mass galaxies also exhibit high stellar mass \textit{fractions} \citep[$f_\star = M_\star/M_{200}$, see e.g. Fig. 8 of][]{S15}, mitigating the influence of the inner halo. Nevertheless, the trend is qualitatively similar for galaxies of all masses, and the corollary is thus that the anisotropy of a galaxy's stellar velocity dispersion is in part governed by the morphology of its inner dark matter halo, with flattened haloes inducing greater anisotropy. The intrinsic morphology of dark matter haloes is likely governed by a combination of their formation time and their intrinsic spin; \citet{Allgood_et_al_06} note that earlier forming haloes (at fixed mass) are systematically more spherical, \citet{Bett_et_al_07} show that intrinsically flatter haloes exhibit a small but systematic offset to greater spin values, and \citet{Jeeson_Daniel_et_al_11} found that formation time (or concentration) and spin are the first two principal components governing dark matter halo structure. These properties emerge simply from the distribution of fluctuations in the initial conditions of the simulations.

\section{Summary and discussion}
\label{sec:summary}

We have performed a quantitative comparison between diagnostics for the morphology and internal kinematics of the stellar component of galaxies in the EAGLE suite of cosmological simulations, and investigated the origin of scatter in this relation. We consider 4155 present-day central galaxies with stellar masses $M_\star > 10^{9.5}\,\Msun$, and in later analyses focus on the subset of 2703 spheroidal or oblate galaxies whose structural and kinematic axes are well-aligned. Our results can be summarized as follows: 

\begin{enumerate}

\item Comparison of five diagnostic quantities frequently used to describe the internal kinematics of the stellar particles comprising simulated galaxies, namely the disc-to-total stellar mass fraction, $D/T$, computed assuming the bulge mass is equal to twice the mass of counter-rotating stellar particles; the fraction of kinetic energy in ordered co-rotation, $\kappa_{\rm co}$; the mass-weighted spin parameter, $\lambda_\star$; the median orbital ellipticity, $\overline{\xi}_i$; and the ratio of the rotational and dispersion velocities, $v_{\rm rot}/\sigma_0$, reveals that they are strongly correlated. This indicates that such descriptors can in general be used interchangeably (Fig. \ref{fig:kinem_metrics}).

\item Modelling EAGLE galaxies as ellipsoids described by the flattening ($\epsilon$) and triaxiality ($T$) parameters (eq. \ref{eq:shapeparameters}) provides a quantitative description of their stellar morphology that is consistent with their qualitative visual appearance. The sample exhibits a diversity of morphologies, including spheroidal, oblate and prolate galaxies. The majority of the sample are oblate ($T \lesssim 0.3$) and flattened ($\epsilon \gtrsim 0.4$), characteristics of ``disky'' galaxies (Fig. \ref{fig:morphology}).

\item The distribution of the shape parameters in the ($u^\star$-$r^\star$) colour - stellar mass plane shows that star-forming central galaxies (comprising the blue cloud) are typically flattened, oblate (rotation-supported) galaxies. The red sequence (of central galaxies) is comprised primarily of spheroidal galaxies at low masses, whilst the more massive regime is dominated by flattened, prolate (dispersion-supported) galaxies. Since both the blue cloud and the red sequence are populated by flattened galaxies, a threshold in $\epsilon$ does not separate the two populations as effectively as as a kinematic criterion such as the $\kappa_{\rm co}=0.4$ threshold advocated by \citet{Correa_et_al_17} (Fig. \ref{fig:colour_mag}). We show that a diagnostic constructed from both shape parameters, $\alpha_{\rm }=(\epsilon^2 + 1 -T)/2$, designed to separate spheres and prolate spheroids from oblate spheroids, is able to separate the two populations with similar efficacy to a kinematic criterion. However we note that $T$ is not readily-accessible from observations, making kinematical diagnostics such as $v_{\rm rot}/\sigma_0$ preferable.

\item Examination of the internal kinematics (quantified via $v_{\rm rot}/\sigma_0$) as a function of morphology (quantified via the flattening, $\epsilon$) reveals a correlation between the two: as expected from dynamical considerations, rotationally-supported galaxies tend to be flatter than dispersion-supported counterparts. However, for all but the most rotationally-supported galaxies, there is significant scatter so that the population of galaxies at fixed $v_{\rm rot}/\sigma_0$ exhibits a broad range of morphologies. The most-massive galaxies ($M_* \gtrsim 10^{10.5}\Msun$) tend to populate the high-$\epsilon$ regime, being either rotationally-supported discs or prolate spheroids. Excision of galaxies with prolate morphology and/or mis-aligned structural and kinematic axes enables analysis of the morpho-kinematics of the remaining subsample with the tensor virial theorem (Fig. \ref{fig:m_k_samples}).

\item The tensor virial theorem (Appendix \ref{sec:virial}) indicates that the flattening of a collisionless system, at fixed $v_{\rm rot}/\sigma_0$, is governed by the anisotropy of its velocity dispersion, $\delta$ (eq. \ref{eq:anisotropy_param}). This prediction is borne out, in a quantitative sense, by the simulated galaxies. At any $v_{\rm rot}/\sigma_0$, more flattened oblate galaxies exhibit greater $\delta$, for all galaxy masses (Fig. \ref{fig:m_k_anisotropy}).

\item A similar trend to that shown in Fig. \ref{fig:m_k_anisotropy} is seen if one correlates $\epsilon$ at fixed $v_{\rm rot}/\sigma_0$ with the flattening of the inner ($< 30 \pkpc$) dark matter halo, $\epsilon_{\rm dm}$. This suggests that a galaxy's morphology is influenced in part by the morphology of its host halo, which is a proxy for the structure of the potential in the region traced by stellar particles. We verify that this is an intrinsic (rather than induced) correlation by measuring $\epsilon_{\rm dm}$ in both the Reference EAGLE simulation (denoting this quantity $\epsilon_{\rm dm}^{\rm Ref}$) and in a simulation considering only collisionless dynamics starting from identical initial conditions ($\epsilon_{\rm dm}^{\rm DMO}$), finding similar trends in both cases (Fig. \ref{fig:m_k_eps_for_dm}).

\item The anisotropy $\delta$ correlates with the flattening of the inner dark matter halo, regardless of whether one considers the flattening of the halo in the Reference simulation, $\epsilon_{\rm dm}^{\rm Ref}$, or its counterpart in the dark-matter-only simulation, $\epsilon_{\rm dm}^{\rm DMO}$ (Fig. \ref{fig:eps_dm_correlations}).

\end{enumerate}

We point out that the link we have established between the shapes of galaxies and the flattening of the inner dark matter halo differs in a fundamental way from previous work on the {\it alignments} of galaxies with surrounding matter.  Indeed, it is well established both theoretically and observationally that galaxies tend to preferentially align themselves with the (dark matter-dominated) large-scale potential (e.g., \citealt{Deason_et_al_11,Velliscig_et_al_15a,Velliscig_et_al_15b,Welker_et_al_17,Welker_et_al_18}).  This leads to so-called ``intrinsic alignments'' of neighbouring galaxies, which acts as a major source of error in measurements of cosmic shear (e.g., \citealt{Hirata_and_Seljak_04,Bridle_and_King_07}).  Our work demonstrates that, not only do galaxies tend to align themselves in a preferential way, but their actual shapes are also determined, to an extent, by the shape of the (local) dark matter potential well.

Our finding that the anisotropy of the stellar velocity dispersion of galaxies correlates with the intrinsic morphology of their inner dark matter haloes is intriguing. The finding that the correlation persists when using the morphology of the inner halo in the corresponding dark matter only simulation is indicative of a causal connection \citep[see also][]{Zavala_et_al_16}. In such a scenario, the formation of a dark matter halo whose inner mass distribution is intrinsically flattened (in the absence of the dissipative physics of galaxy formation) will foster the formation of a galaxy whose stellar velocity dispersion is preferentially expressed in the plane orthogonal to the axis of rotation. As predicted by the tensor virial theorem, this anisotropy fosters the formation of a galaxy that is flatter than typical for galaxies with similar internal $v_{\rm rot}/\sigma_0$. 

The relationship between $\delta$ and $\epsilon_{\rm dm}$ revealed by EAGLE is, in principle, testable with observations. If one stacks galaxies of similar flattening in bins of $v_{\rm rot}/\sigma_0$, and measures the flattening of the total matter distribution (e.g. with weak gravitational lensing), the simulations indicate that one should expect the latter to be systematically greater for galaxies of lower $v_{\rm rot}/\sigma_0$. We note that the overlap of the SDSS-IV/MaNGA integral field survey with the deep imaging fields of the Hyper Suprime-Cam (HSC) survey offers a potential means by which this might be achieved.

\section*{Acknowledgements}  
\label{sec:acknowledgements}

The authors thank the anonymous referee for an insightful and constructive report that improved our study, Camila Correa and James Trayford for helpful discussions, and John Helly for assistance adding new data to the public EAGLE database. ACRT and RAC gratefully acknowledge the support of the Royal Society via a doctoral studentship and a University Research Fellowship, respectively. MS is supported by the Netherlands Organisation for Scientific Research (NWO) through VENI grant 639.041.749. CDPL is funded by the ARC Centre of Excellence for All Sky Astrophysics in 3 Dimensions (ASTRO 3D), through project number CE170100013. This project has received funding from the European Research Council (ERC) under the European Union's Horizon 2020 research and innovation programme (grant agreement number 769130). This work made use of high performance computing facilities at Liverpool John Moores University, partly funded by the Royal Society and LJMU's Faculty of Engineering and Technology, and the DiRAC Data Centric system at Durham University, operated by the Institute for Computational Cosmology on behalf of the STFC DiRAC HPC Facility (www.dirac.ac.uk). This equipment was funded by BIS National E-infrastructure capital grant ST/K00042X/1, STFC capital grant ST/H008519/1, and STFC DiRAC Operations grant ST/K003267/1 and Durham University. DiRAC is part of the National E-Infrastructure. 




\bibliographystyle{mnras}
\bibliography{bibliography} 



\begin{appendix}

\section{Numerical convergence tests}
\label{sec:convergence}

Here we briefly assess the convergence of the kinematics (using the rotation-to-dispersion velocity ratio, $v_{\rm rot}/\sigma_0$) and morphology (using the flattening, $\epsilon$) of EAGLE galaxies with respect to resolution. We consider the `weak convergence' test\footnote{The concept of strong and weak convergence testing was introduced by \citet{S15}.}, comparing the properties of galaxies formed by the Reference model at standard resolution with those of galaxies formed by the Recalibrated model at high-resolution. 

\begin{figure}
\centering
\adjustbox{trim={.00\width} {.01\height} {.05\width} {.08\height},clip}%
  {\includegraphics[width=0.5\textwidth,keepaspectratio]{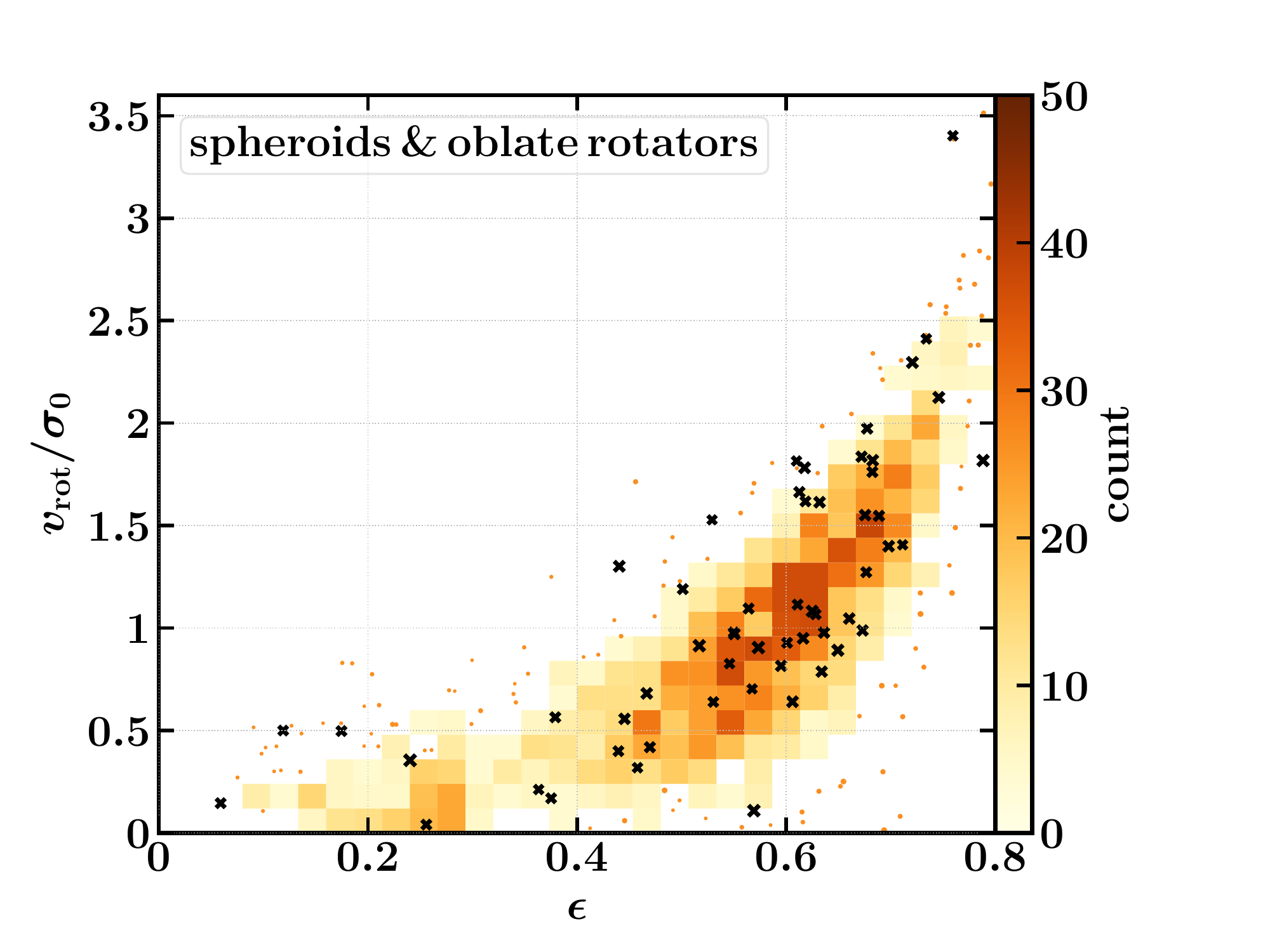}} 
 \caption{The relationship between $v_{\rm rot}/\sigma_0$ and flattening, $\epsilon$, for galaxies with $M_\star > 10^{9.5}\Msun$ and which are i) spheroidal ($\epsilon < 0.3$) or ii) oblate ($T<1/3$) galaxies and have their morphological and rotational axes aligned to within 10 degrees. The 2-dimensional histogram and orange dots represent the 2703 galaxies of Ref-L100N1504 that satisfy these criteria (as per the bottom panel of Fig. \ref{fig:m_k_samples}), whilst black crosses denote the corresponding 57 galaxies from Recal-L025N0752. Visual inspection suggests that the weak convergence behaviour of $v_{\rm rot}/\sigma_0$ and $\epsilon$ is good, an impression corroborated by the two-sample Kolmogorov-Smirnoff test in each case.}
\label{fig:m_k_convergence}
\end{figure}

In Fig. \ref{fig:m_k_convergence} we show as a 2-dimensional histogram in $v_{\rm rot}/\sigma_0-\epsilon$ space the distribution of the sub-set of 2703 galaxies from our main sample in the Ref-L100N1504 simulation that are i) spheroidal ($\epsilon < 0.3$) or ii) oblate ($T<1/3$) galaxies and have their morphological and rotational axes aligned to within 10 degrees. The overlaid crosses denote the 57 galaxies that satisfy the same selection criteria in the high-resolution Recal-L025N0752 simulation. Inspection of the figure suggests that the distributions of $v_{\rm rot}/\sigma_0$ and $\epsilon$ from the two simulations are similar. We obtain a quantitative measure of this similarity using the two-sample Kolmogorov-Smirnoff test, to assess the hypothesis that the values of $v_{\rm rot}/\sigma_0$ and $\epsilon$ for galaxies in the Ref-L100N1504 and Recal-L025N0752 simulations share similar distributions. In both cases the test indicates that this hypothesis cannot be rejected; for $v_{\rm rot}/\sigma_0$ we obtain ($D = 0.11$, $p=0.23$) whilst for $\epsilon$ we obtain ($D = 0.11$, $p=0.20$), where $D$ is the Kolmogorov-Smirnoff statistic and $p$ is the two-tailed probability value. These findings are unchanged if instead we compare Recal-L025N0752 to Ref-L025N0376, to control against box-size effects.

\section{The influence of velocity dispersion anisotropy from the tensor virial theorem}
\label{sec:virial}

\citet{Binney_78} and \citet{Binney_and_Tremaine_87} show that the tensor formulation of the virial theorem provides an analytical framework with which one can relate the morphology of oblate spheroids to their internal dynamics. Starting from
\begin{equation}
2 K_{i j} + W_{i j} = 0,
\label{eq:virialtensor}
\end{equation}
where $K_{i j}$ and $W_{i j}$ represent the kinetic and potential energy tensors, the former can be split into components denoting the streaming motion, $T_{i j}$, and the random motion $\Pi_{i j}$:
\begin{equation}
K_{i j} = T_{i j} + \frac{1}{2} \Pi_{i j}.
\label{eq:kinetictensor}
\end{equation}
For an oblate system rotating about its short axis, we can use the notation $\left ( V_{\rm rot} , \sigma_0 , \delta \right )$ described in Section \ref{sec:kin_diag}, yielding:
\begin{equation}
T_{i j} = \frac{1}{2} M_\star V_{\rm rot}^2 \begin{pmatrix}
1/2 & 0 & 0 \\
0 & 1/2 & 0 \\
0 & 0 & 0
\end{pmatrix} ; 
\Pi_{i j} = M_\star \sigma_0^2 \begin{pmatrix}
1 & 0 & 0 \\
0 & 1 & 0 \\
0 & 0 & 1-\delta
\end{pmatrix}
\label{eq:orderrandom}
\end{equation}
These terms can be related to Cartesian components of the potential energy:
\begin{equation}
\begin{cases}
& \frac{1}{2} M_\star V_{\rm rot}^2 + M_\star \sigma_0^2 = -W_{x x}\\
& M_\star (1-\delta) \sigma_0^2 = -W_{z z}
\end{cases}
\Rightarrow \frac{V_{\rm rot}^2}{\sigma_0^2} = 2 (1-\delta) \frac{W_{x x}}{W_{z z}} - 2
\label{eq:anisotropy}
\end{equation}
where $W_{x x}/W_{z z}$ is related to the flattening parameter, $\epsilon = 1 - c/a$. For convenience let us define the sphericity shape parameter $s=1-\epsilon$ and the term $e$, such that $s \equiv \sqrt{1-e^2}$. From \citet[][see also \citealt{Roberts_1962,Binney_78}]{Binney_and_Tremaine_87}, we have for an oblate body that:
\begin{equation}
\begin{aligned}
&\binom{W_{x x}}{W_{z z}} = - 2 \pi^2 G \frac{c}{a^3} \mathcal{S} \, \binom{A a^2}{C c^2}, \text{  (}\mathcal{S}\text{ being a constant)}\\
&\rm{with} \, \begin{cases}
& A =  \frac{\sqrt{1-e^2}}{e^2} \left ( \frac{\arcsin e}{e} - \sqrt{1-e^2} \right ) \\
& C =  2 \frac{\sqrt{1-e^2}}{e^2} \left ( \frac{1}{\sqrt{1-e^2}} - \frac{\arcsin e}{e} \right )
\end{cases}\\
&\rm{thus} \, \begin{cases}
& A =  \frac{s}{1-s^2} \left ( \frac{\arccos s}{\sqrt{1-s^2}} - s \right ) \\
& C =  2 \frac{s}{1-s^2} \left ( \frac{1}{s} - \frac{\arccos s}{\sqrt{1-s^2}} \right )
\end{cases}\\
& \Rightarrow \frac{W_{x x}}{W_{z z}} = \frac{s^{-2}}{2} \frac{\arccos s - s \sqrt{1-s^2}}{s^{-1} \sqrt{1-s^2} - \arccos s}
\end{aligned}
\label{eq:potentialtensor}
\end{equation}
The dashed tracks overlaid on the left-hand panel of Fig. \ref{fig:m_k_anisotropy} show the relationship between $v_{\rm rot}/\sigma_0$ and $\epsilon$ for $\delta = \left \{ 0,0.1,0.2,0.3,0.4,0.5,0.6 \right \}$, derived using eq. (\ref{eq:anisotropy}) and (\ref{eq:potentialtensor}).

\section{Public release of morphological and kinematical diagnostics}
\label{sec:public}

We have added the values of the particle-based morphological and kinematical diagnostics computed here to the public EAGLE database\footnote{http://galaxy-catalogue.dur.ac.uk}. We thus extend the database with an additional table for each simulation, \texttt{[runname]\_morphokinematics}, where e.g. \texttt{[runname]} is \texttt{RefL0100N1504} for the largest-volume EAGLE simulation. The table contains the shape parameters $\epsilon$ (\texttt{Ellipticity}), $\epsilon_{\rm dm}$ (\texttt{DMEllipticity}) and $T$ (\texttt{Triaxiality}), where $\epsilon_{\rm dm}$ is the value computed for each halo in the hydrodynamical simulation (i.e. not its matched counterpart in the equivalent DMONLY simulation). The following kinematical diagnostics are also included: $D/T$ (\texttt{DiscToTotal}), $\overline{\xi_i}$ (\texttt{MedOrbitCircu}), $\kappa_{\rm co}$ (\texttt{KappaCoRot}), $v_{\rm rot}/\sigma_0$ (\texttt{RotToDispRatio}) and $\delta$ (\texttt{DispAnisotropy}). We do not include values for $\lambda_\star$ since these values were derived from pixel maps and are hence sensitive to choices concerning the construction of the maps. We provide measurements for all 28 snapshots, for the following standard-resolution simulations: Ref-L025N0376, Ref-L050N0752, Ref-L100N1504, AGNdT9-L050N0752\footnote{This is the simulation introduced by \citet{S15}, for which the parameters governing gas accretion onto BHs were recalibrated at the same time as the AGN heating temperature, to maintain an accurate reproduction of the $z=0$ galaxy stellar mass function. It should not be confused with the simulation of the same name presented by \citet{C15}, for which only the AGN heating temperature was varied.}, and the following high-resolution simulations: Ref-L025N0752, Recal-L025N0752.

\citet{McAlpine_et_al_16} describe in detail how to access and query the public EAGLE database, so below we present only a brief example SQL query demonstrating how to retrieve from the database the three morphological and five kinematical diagnostics added here, for sufficiently well-resolved present-day galaxies in the Ref-L100N1504 simulation. We note that the first field of each new table is \texttt{GalaxyID}, an integer that uniquely identifies a galaxy within a particular simulation. This number is consistent with those used in the existing parts of the database, including the extension by \citet{Camps_et_al_17}, so the new tables can be joined with any other table in the database.

\begin{lstlisting}[
           language=SQL,
           showspaces=false,
           basicstyle=\ttfamily,
           numbers=left,
           numberstyle=\tiny,
           commentstyle=\color{gray}
        ]
SELECT 
	MK.Ellipticity AS epsilon,
	MK.Triaxiality AS T,
	MK.DMEllipticity AS epsilon_dm,
	MK.DiscToTotal AS DTratio,
	MK.KappaCoRot AS kappa_co,
	MK.MedOrbitCircu AS orbital,
	MK.RotToDispRatio AS vrotsigratio,
	MK.DispAnisotropy AS delta
FROM
	RefL0100N1504_SubHalo AS SH,
	RefL0100N1504_Aperture AS AP,
	RefL0100N1504_MorphoKinem AS MK
WHERE
	SH.GalaxyID = AP.GalaxyID AND
	SH.GalaxyID = MK.GalaxyID AND
	SH.SubGroupNumber = 0 AND
	AP.ApertureSize = 30 AND
	AP.Mass_Star > 3.16E9 AND
	SH.SnapNum = 28
ORDER BY
	SH.GroupNumber
\end{lstlisting}

\end{appendix}


\bsp	
\label{lastpage}

\end{document}